\documentclass[pdflatex,sn-mathphys-num]{sn-jnl}

\usepackage{graphicx}%
\usepackage{multirow}%
\usepackage{amsmath,amssymb,amsfonts}%
\usepackage{amsthm}%
\usepackage{mathrsfs}%
\usepackage[title]{appendix}%
\usepackage{xcolor}%
\usepackage{textcomp}%
\usepackage{manyfoot}%
\usepackage{booktabs}%
\usepackage{algorithm}%
\usepackage{algorithmicx}%
\usepackage{algpseudocode}%
\usepackage{listings}%
\usepackage{bbold}
\usepackage{slashed}
\usepackage{subfigure}
\usepackage{braket,bm}
\usepackage{cancel}
\usepackage{dsfont}
\usepackage{tikz}
\usepackage{array}
\usepackage{overpic}
\usepackage{float}
\usepackage{threeparttable}
\usepackage{orcidlink}

\graphicspath{{figs/}}

\newcommand{\md}{\mathrm{d}}

\begin{document}

\title[]{Gravitational form factors of pions, kaons and nucleons from dispersion relations}


\author[1]{\fnm{Xiong-Hui} \sur{Cao}
\orcidlink{0000-0003-1365-7178}
}\email{xhcao@itp.ac.cn}

\author[1,2]{\fnm{Feng-Kun} \sur{Guo}
\orcidlink{0000-0002-2919-2064}
}\email{fkguo@itp.ac.cn}

\author[3]{\fnm{Qu-Zhi} \sur{Li}
\orcidlink{0009-0001-2640-1174}
}\email{liquzhi@scu.edu.cn}

\author[4]{\fnm{Bo-Wen} \sur{Wu}}\email{wubowen@hnu.edu.cn}

\author[4,5,1]{\fnm{De-Liang} \sur{Yao}
\orcidlink{0000-0002-5524-2117}
}\email{yaodeliang@hnu.edu.cn}

\affil[1]{\orgdiv{Institute of Theoretical Physics}, \orgname{Chinese Academy of Sciences}, \orgaddress{\city{Beijing}, \postcode{100190}, \country{China}}}

\affil[2]{\orgdiv{School of Physical Sciences}, \orgname{University of Chinese Academy of Sciences}, \orgaddress{\city{Beijing}, \postcode{100190}, \country{China}}}

\affil[3]{\orgdiv{Institute for Particle and Nuclear Physics, College of Physics}, \orgname{Sichuan University}, \orgaddress{\city{Chengdu}, \state{Sichuan} \postcode{610065}, \country{China}}}

\affil[4]{\orgdiv{School of Physics and Electronics}, \orgname{Hunan University}, \orgaddress{\city{Changsha}, \state{Hunan} \postcode{410082}, \country{China}}}

\affil[5]{\orgdiv{Hunan Provincial Key Laboratory of High-Energy Scale Physics and Applications}, \orgname{Hunan University}, \orgaddress{\city{Changsha}, \state{Hunan} \postcode{410082}, \country{China}}}


\abstract{The gravitational form factors of pions, kaons and the nucleons are investigated by employing modern dispersive techniques and chiral perturbation theory. 
We determine the gravitational form factors of pions and kaons, extending our analysis to explore the pion mass dependence of these form factors at several unphysical pion masses up to 391~MeV, for which lattice results exist for the meson-meson scattering phase shifts.
We also review our analysis on the nucleon gravitational form factors at the physical pion mass, and then systematically calculate various three-dimensional spatial and two-dimensional transverse density distributions for the nucleons. These results provide new insights into the mass distribution inside nucleons. 
As a by-product, we match our dispersion relation results and those obtained from chiral perturbation theory with external gravitational source at the next-to-next-to-leading order, yielding values for the low-energy constants $c_8=-4.28_{-0.38}^{+0.37} ~\mathrm{GeV}^{-1}$ and $c_9=-0.68_{-0.05}^{+0.06} ~\mathrm{GeV}^{-1}$. 
These results offer a robust benchmark for future experimental and theoretical studies.}



\maketitle

\tableofcontents

\clearpage

\section{Introduction}\label{sec1}

The exploration of pions, kaons, and nucleons plays a pivotal role in both theoretical and experimental physics. 
Pions and kaons are recognized as pseudo-Nambu-Goldstone bosons (pNGBs), which emerge from the spontaneous breaking of chiral symmetry in quantum chromodynamics (QCD). 
Through chiral dynamics, they provide unique insights into the nonperturbative dynamics of QCD and serve as critical probes of hadronic structure and interactions at low energies.
The internal structures of pNGBs and nucleons are crucial for our understanding of visible matter in the universe. 
In recent years, gravitational form factors (GFFs), defined by the hadronic matrix elements of the energy-momentum tensor (EMT), have attracted intensive interest. 
Analogous to electromagnetic form factors, which encode information about charge distribution within particles~\cite{Thomas:2001kw}, GFFs encode various internal properties such as mass, energy, angular momentum (AM), and internal force distributions~\cite{Polyakov:2002yz, Polyakov:2018zvc}.

In practice, the nucleon GFFs are experimentally accessible because they are weighted integrals over the generalized parton distribution (GPD) functions, which can be accessed from hard exclusive processes such as deeply virtual Compton scattering~\cite{Ji:1996ek, Radyushkin:1997ki} and hard exclusive meson production~\cite{Collins:1996fb}. 
Experimental measurements of GFFs in hard QCD processes have already become available for both the pion~\cite{Kumano:2017lhr} and the nucleon~\cite{Burkert:2018bqq}. 
A detailed exploration of experimental methodologies for extracting GFFs from hard exclusive processes can be found in Refs.~\cite{Kumericki:2019ddg, Dutrieux:2021nlz, Burkert:2023wzr}. 
In addition, hadronic EMT matrix elements also serve as crucial inputs for studies of hadronic decays of heavy quarkonia~\cite{Novikov:1980fa, Voloshin:1980zf, Voloshin:1982eb}, hadronic decays of light Higgs-like scalars~\cite{Donoghue:1990xh, Monin:2018lee, Winkler:2018qyg, Blackstone:2024ouf}, semileptonic $\tau$ decays~\cite{Celis:2013xja}, and even investigations of hidden-charm pentaquarks~\cite{Eides:2015dtr, Perevalova:2016dln}.
Meanwhile, there has been considerable interest in employing near-threshold heavy quarkonium photoproduction to constrain gluonic GFFs of the nucleon~\cite{Kharzeev:1995ij, Hatta:2018ina, Wang:2021dis, Mamo:2022eui, Guo:2023pqw}. However, the validity of this approach remains under debate~\cite{Du:2020bqj,Xu:2021mju, JointPhysicsAnalysisCenter:2023qgg}.

On the theoretical side, various models have been employed to explore the GFFs of pNGBs~\cite{Broniowski:2008hx,Hudson:2017xug, Freese:2019bhb, Aliev:2020aih, Krutov:2022zgg, Xu:2023izo, Li:2023izn, Fujii:2024rqd, Broniowski:2024oyk, Liu:2024vkj, Liu:2024jno, Liu:2025vfe} and nucleons~\cite{Pasquini:2014vua, Hudson:2017oul, Neubelt:2019sou, Anikin:2019kwi, Azizi:2019ytx, Kim:2020nug, Mamo:2022eui, Choudhary:2022den, Guo:2023pqw, Wang:2023fmx, Cao:2023ohj, Yao:2024ixu, Wang:2024abv, Wang:2024fjt, Guo:2025jiz, Dehghan:2025ncw, Broniowski:2025ctl, Goharipour:2025lep, Sain:2025kup, Deng:2025fpq, Nair:2025sfr}. 
However, these model-dependent results are typically subject to uncertainties that are difficult to control. 
Lattice QCD (LQCD) has been used to compute the quark contribution to the GFFs of the pion and proton~\cite{Gockeler:2003jfa, Brommel:2005ee,Brommel:2007zz, LHPC:2007blg, LHPC:2010jcs, Alexandrou:2013joa, Delmar:2024vxn}, the gluonic contribution to the GFFs for various hadrons~\cite{Shanahan:2018nnv, Shanahan:2018pib}, and the gluonic trace anomaly FFs of the pion and nucleon~\cite{Wang:2024lrm}. 
A notable development is the recent LQCD calculations of the pion and nucleon GFFs~\cite{Hackett:2023nkr, Hackett:2023rif} with a pion mass of $m_\pi=170$ MeV, which is close to the physical pion mass. 
Both the light quark and gluonic contributions are considered therein~\cite{Hackett:2023nkr, Hackett:2023rif}, leading to significant improvement compared to earlier LQCD calculations.
Recently, we have computed the pion and nucleon GFFs using the technique of dispersion relation (DR)~\cite{Cao:2024zlf}, and obtained model-independant results in the sense that the method is data driven based on axiomatic principles such as unitarity, analyticity and crossing symmetry (see, e.g. Ref.~\cite{Yao:2020bxx}, for a recent comprehensive review) and the uncertainties are under control. 

While mass and spin are related to the Casimir operators of the Poincar\'e group, the nucleon $D$-term is associated with the stress component of the EMT, which reveals the mechanical properties of the nucleon (for a different point of view, see Ref.~\cite{Ji:2025gsq}). 
However, it remains poorly understood and is therefore regarded as ``the last global unknown property" of the nucleon~\cite{Polyakov:2002yz, Polyakov:2018zvc}. 
Unlike the case of pions, the value of the nucleon $D$-term is not directly determined by chiral symmetry, but is indirectly influenced by the complex internal dynamics of QCD~\cite{Goeke:2001tz}. 
Based on classical mechanical stability arguments, it is conjectured that the $D$-term should be negative~\cite{Perevalova:2016dln, Lorce:2025oot}. 
This conjecture is supported by various chiral soliton model calculations~\cite{Cebulla:2007ei,Goeke:2007fp,Kim:2012ts,Wakamatsu:2007uc,Jung:2013bya,Kim:2021kum}: $-4 \lesssim D \lesssim-1$ for the nucleon, and also by the result from dispersive analysis in Ref.~\cite{Cao:2024zlf}, $D=-3.38_{-0.35}^{+0.34}$.

The DR method is rooted in the Källén-Lehmann spectral representation. The $t$-channel spectral functions of GFFs are expected to be dominated by intermediate states that can go on shell, such as pairs of pions and kaons in the low-energy region. However, for momentum transfers ranging from 0 to $\sim 2$~GeV, a model-independent and precise calculation of these GFFs as well as the $D$-term at the physical quark mass poses a great challenge. 
On the one hand, the validity region of the chiral perturbation theory (ChPT) is restricted to low energies of at most a few hundred MeV above the threshold. 
In fact, the convergence properties of the chiral expansion are worse than one would naively expect due to the presence of the notorious low-lying scalar resonances. 
On the other hand, perturbative QCD (pQCD) only works in the high-energy region. 
The DR techniques can be used to establish a reliable description of the GFFs of pNGBs and nucleons across the entire energy region.
The dispersive analysis of the pion GFFs was pioneered in Ref.~\cite{Donoghue:1990xh} using low-precision data, and was improved in a recent work~\cite{Broniowski:2024oyk} by combining $\pi\pi$-$K\bar{K}$ scattering~\cite{Celis:2013xja} with fitting to the lattice results~\cite{Hackett:2023nkr}. 
In Ref.~\cite{Cao:2024zlf}, we established a DR framework to describe the GFFs of pNGBs and nucleons, where the pion and kaon GFFs are matched to the next-to-leading order (NLO) ChPT at low momentum transfer.
This work, as a follow-up to Ref.~\cite{Cao:2024zlf}, presents a detailed dispersive analysis of the GFFs of both pNGBs (pions and kaons) and nucleons in a model-independent way, encompassing the pion mass dependence of the pion GFFs, the three-dimensional (3D) spatial and two-dimensional (2D) transverse densities of the nucleon, and the matching of dispersion relations to ChPT. For completeness, the framework established in Ref.~\cite{Cao:2024zlf} will also be reviewed. 

The outline of this paper is as follows. 
In Sec.~\ref{sec.2}, we briefly introduce the EMT and the definition of GFFs. 
In Sec.~\ref{sec.3}, we review our framework in Ref.~\cite{Cao:2024zlf} for analyzing the GFFs of pNGBs by considering the $\pi\pi$ and $K\bar{K}$ intermediate states and the corresponding unitarity relations, matched to the NLO ChPT at low momentum transfer; then we present results at a few unphysical pion masses up to 391~MeV. 
The framework for a comprehensive description of the nucleon GFFs is reviewed in Sec.~\ref{sec.4}. 
In Sec.~\ref{sec.5}, we show our results of various internal spatial distributions and the corresponding radii of the nucleons. 
The matching of the DR representation to the ChPT result for the nucleon will be discussed in Sec.~\ref{sec.6}. 
The results of our work are summarized in Sec.~\ref{sec.7}.
Details on the use of the $z$-expansion method for the extrapolation of the GFFs to asymptotically large momentum transfer are provided in Appendix~\ref{app.z-exp}.

\section{Energy-momentum tensor and gravitational form factors}\label{sec.2}

In this section, we first introduce the generic symmetric EMT in field theory and the EMT of QCD. Then, the definition of GFFs of pNGBs and nucleons is specified. Finally, the general properties of the relevant GFFs are briefly reviewed.

\subsection{Generic symmetric EMT in field theory}

Noether's first theorem relates continuous global symmetry transformations to conserved currents. 
Consequently, the invariance under infinitesimal spacetime transformations $x^\mu \rightarrow x^{\prime \mu}=x^\mu+\varepsilon^\mu$ of a system, described by a Lagrangian $\cal{L}$, leads to the canonical EMT,
\begin{align} 
T_{\text{can}}^{\mu \nu} = \frac{\partial \mathcal{L}}{\partial(\partial_\mu \phi)} \partial^\nu \phi - g^{\mu \nu} \mathcal{L}\ , 
\end{align}
where $\phi$ is a real scalar field. Poincar\'e invariance ensures that the canonical EMT is conserved (i.e., $\partial_\mu T_{\text{can}}^{\mu \nu} = 0$). It is symmetric (i.e., $T_{\text{can}}^{\mu \nu} = T_{\text{can}}^{\nu \mu}$) for spin-0 scalar fields, but asymmetric for higher-spin fields. 
The Belinfante improvement addresses this issue by adding a divergence term $\partial_\lambda X^{\lambda \mu \nu}$ satisfying $X^{\lambda \mu \nu} = -X^{\mu \lambda \nu}$, so that current conservation is preserved while the Lorentz indices become symmetric.

Another commonly used method to derive a symmetric EMT consists of coupling the theory to a weak classical background gravitational field described by a symmetric metric field $g_{\mu \nu}(x)$. 
For instance, the Lagrangian for a real scalar field $\mathcal{L}$ can be generalized to a curved spacetime by replacing $\left(\partial_\mu \phi\right)\left(\partial^\mu \phi\right)$ with $g_{\mu \nu}(x)\left(\partial^\mu \phi\right)\left(\partial^\nu \phi\right)$. 
One obtains the symmetric EMT by varying the action $S_{\text {grav }}=\int \md^4 x \sqrt{|g|} \mathcal{L}$ with respect to the background field according to
\begin{align}
    T_{\mu \nu}=\frac{2}{\sqrt{|g|}} \frac{\delta S_{\mathrm{grav}}}{\delta g^{\mu \nu}}\bigg |_{g_{\mu\nu}=\eta_{\mu\nu}}\ ,
\end{align}
where $g$ denotes the determinant of the metric tensor $g^{\mu\nu}$, and $\eta^{\mu\nu}$ is the metric tensor in flat spacetime. 
A pedagogical description of this method can be found in Appendix~E of Ref.~\cite{Belitsky:2005qn}. 
In this work, we utilize the symmetric EMT (for a discussion of the asymmetric EMT, see Refs.~\cite{Lorce:2017wkb,Freese:2025glz} and references therein).

\subsection{The EMT in QCD}

The symmetric Belinfante-improved total QCD EMT operator is defined as
\begin{align}
    \hat{T}^{\mu \nu}=\sum_q T_q^{\mu \nu}+T_g^{\mu \nu}\ .
\end{align}
The quark and gluon contributions to the total EMT are given by\footnote{In the Faddeev-Popov quantization of gauge theory, the ghost and gauge-fixing terms appear in the Lagrangian, leading to corresponding terms in the QCD EMT. 
However, Becchi-Rouet-Stora-Tyutin symmetry ensures that these terms vanish in physical matrix elements~\cite{Joglekar:1975nu}. 
Therefore, we exclude them from this expression.}
\begin{align}
    \begin{aligned}
    T_q^{\mu \nu}= & \bar{\psi}_q \left(\frac{i}{2}\gamma^{\{\mu}\mathcal{D}^{\nu\}}+\frac{1}{4}g^{\mu\nu}m_q\right)\psi_q\ , \\
    T_g^{\mu \nu}= & F^{A, \mu \eta} F^{A,}{ }_\eta{ }^\nu+\frac{1}{4} g^{\mu \nu} F^{A, \kappa \eta} F^{A,}{ }_{\kappa \eta}\ ,
\end{aligned}
\end{align}
where $\psi_q$ and $\mathcal{A}^A_\mu$ are quark and gluon fields, respectively, $m_q$ is the quark mass of flavor $q$, the superscript $A$ denotes the color index of the gluon field, and the notation $a^{\{\mu} b^{\nu\}}=a^\mu b^\nu+a^\nu b^\mu$ has been used. The covariant derivatives are defined as $\mathcal{D}_\mu=\partial_\mu-i g_s T^A \mathcal{A}_\mu^A$, $F_{\mu \nu}^A=\partial_\mu \mathcal{A}_\nu^A-\partial_\nu \mathcal{A}_\mu^A+g_s f^{A B C} \mathcal{A}_\mu^B \mathcal{A}_\nu^C$, with $g_s$ the strong coupling and $f^{ABC}$ the SU(3) structure constants, and the SU(3) color group generators are normalized as $\operatorname{tr}\left(T^A T^B\right)=$ $\frac{1}{2} \delta^{A B}$. 
The total EMT is conserved; however, the gluon and quark parts of the EMT are not individually conserved. 
Moreover, the total EMT operator is ultraviolet-finite and scale-independent~\cite{Freedman:1974gs}, while the quark or gluon sector of the EMT is ultraviolet-divergent and scale-dependent. 
Classically, massless QCD is invariant under scale transformations of spacetime, 
\begin{align}
    x^\mu\rightarrow \lambda^{-1}x^\mu, \quad 
    \psi_q(x)\rightarrow \lambda^{3/2}\psi_q(\lambda x), \quad 
    \mathcal{A}^A_\mu(x)\rightarrow \lambda \mathcal{A}^A_\mu(\lambda x)
\end{align}
for an arbitrary $\lambda>0$. 
For massless QCD, the associated dilation current $j^\mu=x_\nu \hat{T}^{\mu \nu}$ is conserved at the classical level~\cite{Callan:1970ze}, since it satisfies $\partial_\mu j^\mu=\hat{T}_\mu{ }^\mu$ and $\hat{T}_\mu{ }^\mu=\sum_q m_q \bar{\psi}_q \psi_q=0$ in the chiral limit $m_q=0$. 
However, quantum corrections break the classical scale symmetry due to the trace anomaly in non-Abelian gauge theories~\cite{Collins:1976yq, Nielsen:1977sy},
\begin{align}\label{eq.trace anomaly}
    \hat{T}_\mu{ }^\mu \equiv \frac{\beta(g_s)}{2 g_s} F^{A, \mu \nu} F^{A,}{ }_{\mu \nu}+\left(1+\gamma_m\right) \sum_q m_q \bar{\psi}_q \psi_q\ ,
\end{align}
where $\beta(g_s)$ is the $\beta$-function of QCD, and $\gamma_m$ is the anomalous dimension of the mass operator. 

\subsection{Definition of gravitational form factors}

Let us first define the spacelike GFFs of a spin-0 particle, e.g., the pion. 
We use the covariant normalization of the one-particle state $\left\langle p^{\prime} \mid p\right\rangle=2 p^0(2 \pi)^3 \delta^3\left(\boldsymbol{p}^{\prime}-\boldsymbol{p}\right)$, and introduce the combinations of momenta:\footnote{Many authors also use the average momentum, defined as $\bar{P}^\mu=P^\mu/2=\left(p^{\mu \prime}+p^\mu\right)/2$, instead of the total momentum $P^\mu$.} $P^\mu=p^{\mu \prime}+p^\mu$ and $\Delta^\mu=p^{\mu \prime}-p^\mu$. 
In a theory that is invariant under parity, charge conjugation, and time reversal, the total EMT matrix elements $\left\langle p^{\prime}\left|\hat{T}^{\mu \nu}(0)\right| p\right\rangle$ can be expressed in terms of Lorentz structures constructed from $P^\mu$, $\Delta^\mu$, and $g^{\mu \nu}$. Note that, for particles with spin, one also needs to take the Dirac gamma matrices into account. 
The constraint $\Delta_\mu \left\langle p^{\prime}\left|\hat{T}^{\mu \nu}(0)\right| p\right\rangle = 0$ must be satisfied due to the conservation of total energy and momentum. 
Consequently, only two independent symmetric tensors, $P^\mu P^\nu$ and $\left(\Delta^\mu \Delta^\nu - g^{\mu \nu} \Delta^2\right)$, are possible.
Therefore, a spin-0 pion is characterized by two total GFFs, which are defined as~\cite{Pagels:1966zza,Donoghue:1991qv,Kubis:1999db,Hudson:2017xug,Cotogno:2019vjb}
\begin{align}\label{eq.pi_space}
    \left\langle\pi^a(p^{\prime})\left|\hat{T}^{\mu \nu}(0)\right| \pi^b(p)\right\rangle=\frac{\delta^{ab}}{2}\left[A^\pi(t)P^\mu P^\nu+D^\pi(t)\left(\Delta^\mu \Delta^\nu-t g^{\mu \nu}\right) \right] ,
\end{align}
where $t\equiv \Delta^2<0$ and $a,b=1,2,3$ are isospin indices.
We will work in the exact isospin symmetric limit.
Then, by crossing we obtain the timelike GFFs from Eq.~\eqref{eq.pi_space} as
\begin{align}\label{eq.pi_time}
    \left\langle\pi^a(p^\prime) \pi^b\left(p\right)\left|\hat{T}^{\mu \nu}(0)\right| 0\right\rangle=\frac{\delta^{ab}}{2}\left[A^\pi(t)\Delta^\mu \Delta^\nu +D^\pi(t)\left(P^\mu P^\nu-t g^{\mu \nu}\right) \right] ,
\end{align}
with $t\equiv P^2>0$. 

Likewise, the total GFFs of a spin-$1/2$ nucleon are defined as~\cite{Kobzarev:1962wt,Pagels:1966zza,Ji:1996ek,Cotogno:2019vjb}
\begin{align}\label{eq.nuclon_space}
    & \left\langle N(p^{\prime})\left|\hat{T}^{\mu \nu}(0)\right| N(p)\right\rangle= \nonumber\\
    & \frac{1}{4m_N}\bar{u}(p^{\prime})\left[\hat A(t) P^\mu P^\nu+\hat J(t)\left(i P^{\{\mu} \sigma^{\nu\} \rho} \Delta_\rho\right)+\hat D(t) \left(\Delta^\mu \Delta^\nu-t g^{\mu \nu}\right)\right] u(p)\ ,
\end{align}
where the normalization of the spinors is $\bar{u}(p,s) u(p,s)=2 m_N$. Imposing the Gordon identity $2 m_N \bar{u}^{\prime} \gamma^\alpha u=$ $\bar{u}^{\prime}\left(P^\alpha+i \sigma^{\alpha \rho} \Delta_\rho\right) u$, an alternative decomposition is obtained:
\begin{align}
    & \left\langle N(p^{\prime})\left|\hat{T}^{\mu \nu}(0)\right| N(p)\right\rangle= \nonumber\\
    & \frac{1}{4m_N}\bar{u}(p^{\prime})\left[m_N \hat A(t) \gamma^{\{\mu} P^{\nu\}}+ \frac{\hat B(t)}{2}\left(i P^{\{\mu} \sigma^{\nu\} \rho} \Delta_\rho\right)+\hat D(t) \left(\Delta^\mu \Delta^\nu-t g^{\mu \nu}\right)\right] u(p)\ .
\end{align}
The two representations are equivalent, with the GFFs related by $\hat A(t)+\hat B(t)=2 \hat J(t)$. 
From Eq.~\eqref{eq.nuclon_space}, the nucleon timelike GFFs are defined as
\begin{align}\label{eq.nucleon_time}
    & \left\langle N(p^{\prime})\bar{N}(p)\left|\hat{T}^{\mu \nu}(0)\right| 0\right\rangle= \nonumber\\
    & \frac{1}{4m_N}\bar{u}(p^{\prime})\left[\hat A(t) \Delta^\mu \Delta^\nu+\hat J(t)\left(i \Delta^{\{\mu} \sigma^{\nu\} \rho} P_\rho\right)+\hat D(t) \left(P^\mu P^\nu-tg^{\mu \nu}\right)\right] u(p)\ .
\end{align}
For a dispersive analysis of GFFs, it is often convenient to work in the isospin basis and decompose the GFFs into isoscalar ($s$) and isovector ($v$) components,
\begin{align}
    \hat X&=X^s \mathbb{1}+X^v \tau^3\ ,\quad X=\{A,B,J,D\}\ ,
\end{align}
with $\tau^3$ the third Pauli matrix acting in the isospin space.
The isoscalar and isovector GFFs, $X^s$ and $X^v$, are related to the physical ones, $X^p$ and $X^n$, as
\begin{eqnarray}
    \left\{
    \begin{array}{c}
        X^s=\frac{1}{2}\left(X^p+X^n\right)   \\
        X^v=\frac{1}{2}\left(X^p-X^n\right) 
    \end{array}
    \right. \ ,\quad
     \left\{
    \begin{array}{c}
        X^p=X^s+X^v   \\
        X^n=X^s-X^v 
    \end{array}
    \right. \ .\label{eq.isospinv1}
\end{eqnarray}
Experimental data are usually given in terms of the physical GFFs.

\subsection{Properties of gravitational form factors}

For a physical hadronic scattering process, one always has $t<0$ or $t\geq 4m_\pi^2$, and the point $t=0$ can only be reached through reliable extrapolation.
The GFF $A(t)$, i.e., the term associated with the Lorentz structure $P^\mu P^\nu$, can be defined for a particle of any spin. The extrapolation of $A(t)$ to the kinematical point $t = 0$ yields~\cite{Pagels:1966zza}
\begin{align}\label{eq.A0}
    \lim_{t \rightarrow 0} A(t)=A(0)=1\ .
\end{align}
The normalization in Eq.~\eqref{eq.A0} can be understood by recalling that for $\boldsymbol{p} \rightarrow \mathbf{0}$ and $\boldsymbol{p}^{\prime} \rightarrow \mathbf{0}$, only the 00-component of the EMT remains in Eqs.~\eqref{eq.pi_space} and~\eqref{eq.nuclon_space}. The Hamiltonian of the system can then be obtained as $H=\int \mathrm{d}^3 x \hat{T}_{00}(x)$ and satisfies $H|\boldsymbol{p}\rangle=m|\boldsymbol{p}\rangle$ as $\boldsymbol{p} \rightarrow \mathbf{0}$ with $m$ mass of the particle. 

The GFF $J(t)$ is absent in the spin-0 case but present in the spin-1/2 case. In the limit $t \rightarrow 0$, this GFF satisfies the constraint
\begin{align}\label{eq.J0}
    \lim _{t \rightarrow 0} J(t)=J(0)=\frac{1}{2}\ ,
\end{align}
which is simply the spin of the particle~\cite{Ji:1996ek}. 
The constraints in Eqs.~\eqref{eq.A0} and~\eqref{eq.J0} have been rigorously proven recently in Ref.~\cite{Lowdon:2017idv}.

However, the GFF $D(t)$ at zero-momentum transfer is unconstrained. Its value is referred to as the $D$-term, $D \equiv D(0)$. Any hadron, regardless of its mass and spin, possesses a $D$-term. 
It is instructive to give some simple examples of the value of the $D$-term.
\begin{itemize}
    \item[$1)$] Consider a free spin-0 boson as described by the free Klein-Gordon theory~\cite{Birrell:1982ix},
    \begin{align}
        \mathcal{L}=\sqrt{|g|}\left(\frac{1}{2}g^{\mu\nu}(x)\left(\partial_\mu \phi\right)\left(\partial_\nu \phi\right)-\frac{1}{2} m^2 \phi^2-\frac{1}{2}hR(x)\phi^2\right) ,
        \label{eq:Lspin0_free}
    \end{align}
    where $-h R(x) \phi^2/2$ is a non-minimal coupling term, and $R$ is the Ricci curvature scalar. In fact, this term does not violate any symmetry of the free scalar field and is also a renormalizable dimension-4 operator.
    Therefore, the GFF $D(t)$ and the $D$-term are $D(t)=D=-1+4h$, whose value is unknown as $h$ is free. If we make a trivial choice $h=0$ (also called the minimally coupled case), it results in $D=-1$~\cite{Pagels:1966zza, Hudson:2017xug}. 
    However, the choice of $h$ is not unique and  actually relies on additional symmetry and renormalization requirements of the theory.
    \item[$2)$] The weakly interacting $\phi^4$ theory is defined by~\cite{Birrell:1982ix}
    \begin{align}
        \mathcal{L}=\sqrt{|g|}\left(\frac{1}{2}g^{\mu\nu}(x)\left(\partial_\mu \phi\right)\left(\partial_\nu \phi\right)-\frac{1}{2} m^2 \phi^2-\frac{\lambda}{4!} \phi^4-\frac{1}{2}hR(x)\phi^2\right) .
    \end{align}
    Due to the presence of the four-particle $\phi^4$ interaction, one-loop renormalizability requires $h=1/6$\footnote{This is also called the conformally coupled case~\cite{Callan:1970ze}, which means that if $m=0$ the action and hence the field equations are invariant under conformal transformations.} and then $D=-1+4\times 1/6=-1/3$~\cite{Callan:1970ze}; see also Refs.~\cite{Hudson:2017xug,Maynard:2024wyi}. 
    To be more specific, the choice $h=1/6$ removes UV divergences of $D(t)$ up to three loops in dimensional regularization in $\phi^4$ theory, although the value of $h$ needs to be changed beyond three loops~\cite{Collins:1976vm}.
    \item[$3)$] For a free Dirac fermion, say the nucleon, it is described by the Lagrangian~\cite{Birrell:1982ix}
    \begin{align}
        \mathcal{L}=\sqrt{|g|} \left( \frac{i}{2} \left( \bar{\psi} \gamma^\mu \nabla_\mu \psi - \nabla_\mu \bar{\psi} \gamma^\mu \psi \right) - m \bar{\psi} \psi \right) ,
    \end{align}
    where the covariant derivative acting on the nucleon field has the form 
    \begin{align*}
    \nabla_\mu \psi=\partial_\mu \psi+\frac{i}{2} \omega_\mu^{a b} \sigma_{a b} \psi\ , \quad \nabla_\mu \bar{\psi}=\partial_\mu \bar{\psi}-\frac{i}{2} \bar{\psi} \sigma_{a b} \omega_\mu^{a b}\ ,
    \end{align*}
    with $\sigma_{ab}=i[\gamma_a, \gamma_b]/2$ and
    \begin{align*}
    \omega_\mu^{a b}=-g^{\nu \lambda} e_\lambda^a\left(\partial_\mu e_\nu^b-e_\sigma^b \Gamma_{\mu \nu}^\sigma\right) ,\quad \Gamma_{\alpha \beta}^\lambda=\frac{1}{2} g^{\lambda \sigma}\left(\partial_\alpha g_{\beta \sigma}+\partial_\beta g_{\alpha \sigma}-\partial_\sigma g_{\alpha \beta}\right) .
    \end{align*}
    The vielbein fields $e^a_\mu$ satisfy the following identities:
    \begin{align*}
    e_\mu^a e_\nu^b \eta_{a b}=g_{\mu \nu}\ ,\quad e_a^\mu e_b^\nu \eta^{a b}=g^{\mu \nu}\ ,\quad e_\mu^a e_\nu^b g^{\mu \nu}=\eta^{a b}\ ,\quad e_a^\mu e_b^\nu g_{\mu \nu}=\eta_{a b}\ ,
    \end{align*}
    with $\eta_{ab}$ the flat Minkowski space matric tensor.
    The above Lagrangian leads to a vanishing $D$-term, i.e., $D=D(t)=0$~\cite{Hudson:2017oul}.
    \item[$4)$] Chiral symmetry uniquely determines the interactions of pNGBs in the chiral limit. 
    The standard ``improvement'' of scalar field theory, which amounts to supplementing a term proportional to $R\phi^2$ in Eq.~\eqref{eq:Lspin0_free}, cannot apply here because such a term breaks chiral symmetry~\cite{Voloshin:1982eb,Leutwyler:1989tn} as it is of a nonderivative form. 
    Furthermore, ChPT, as an effective field theory of low-energy QCD, is non-renormalizable in the sense of Bogoliubov-Parasiuk-Hepp-Zimmermann (BPHZ) renormalization scheme and can only be renormalized order by order.
    In Refs.~\cite{Novikov:1980fa,Voloshin:1980zf}, the following low-energy theorem was derived using current algebra,
    \begin{align}
        D^i=-1+\mathcal{O}(p^2), \quad i=\pi, K, \eta,
    \end{align}
    that is, one has $D^i=-1$ in the chiral limit. 
    The chiral corrections of the GFFs, $A^i(t)$ and $D^i(t)$, were studied in SU(3) ChPT up to $\mathcal{O}(p^4)$ in Ref.~\cite{Donoghue:1991qv}.
    We quote here only the results for the $D$-terms:
    \begin{align}
        & D^\pi=-1+16 L^r \frac{m_\pi^2}{F_\pi^2}+\frac{m_\pi^2}{F_\pi^2} I_\pi-\frac{m_\pi^2}{3 F_\pi^2} I_\eta+\mathcal{O}(p^4)\ , \\
        & D^K=-1+16 L^r \frac{m_K^2}{F_\pi^2}+\frac{2 m_K^2}{3 F_\pi^2} I_\eta+\mathcal{O}(p^4)\ , \\
        & D^\eta=-1+16 L^r \frac{m_\eta^2}{F_\pi^2}-\frac{m_\pi^2}{F_\pi^2} I_\pi+\frac{8 m_K^2}{3 F_\pi^2} I_K+\frac{4 m_\eta^2-m_\pi^2}{3 F_\pi^2} I_\eta+\mathcal{O}(p^4)\ ,
    \end{align}
    with
    \begin{align}
        L^r=L_{11}^r(\mu)-L_{13}^r(\mu), \quad I_i=\frac{1}{48 \pi^2}\left(\ln \left(\frac{\mu^2}{m_i^2}\right)-1\right) .
    \end{align}
    In the $D$-term, the scale dependence in the low-energy constants (LECs) $L_{11}^r(\mu)$ and $L_{13}^r(\mu)$ cancels out that in the chiral logarithms $I_i$.
    In Ref.~\cite{Donoghue:1991qv}, the LECs were estimated using DR techniques and the scalar meson dominance model as\footnote{Various chiral models can be used to estimate the curvature-induced LECs $L_{11},L_{12}$ and $L_{13}$; see Refs.~\cite{Andrianov:1998fr,Megias:2004uj,Megias:2005fj}. Recently, an extraction of the above LECs from lattice QCD calculations has been carried out in Ref.~\cite{RuizArriola:2024udm}.}
    \begin{align}
        L_{11}^r(1~\mathrm{GeV})=(1.4-1.6) \times 10^{-3}\ ,\quad L_{13}^r(1~\mathrm{GeV})=(0.9-1.1) \times 10^{-3}\ ,
    \end{align}
    The above LECs yield
    \begin{align}
        D^\pi=-0.97 \pm 0.01\ ,\quad D^K=-0.76 \pm 0.09\ ,\quad D^\eta=-0.65 \pm 0.12\ ,
    \end{align}
    where the uncertainties are propagated from the errors of the LECs. 
    Here we use the physical pion decay constant, $F_\pi=92.1~\mathrm{MeV}$~\cite{ParticleDataGroup:2024cfk}.
\end{itemize}

\section{Dispersive representation of pion and kaon gravitational form factors}\label{sec.3}

\subsection{Unitarity and spectral function}

\begin{figure}[t]
    \centering
    \includegraphics[width=.68\textwidth,angle=-0]{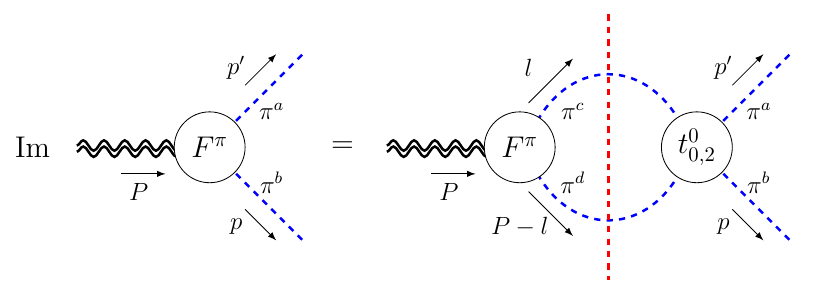}
    \caption{Elastic unitarity relation for the pion GFFs $F^\pi=\{A^\pi,D^\pi\}$~\cite{Cao:2024zlf}. The blue dashed lines denote the pions, the double wiggly lines represent the external QCD EMT current, and the dashed red vertical line indicates that the intermediate pions are on-shell.}\label{fig.cut_pi}
\end{figure}

Here we give a brief review of the dispersive representation of the pion and kaon GFFs presented in Ref.~\cite{Cao:2024zlf}.
The imaginary parts of the pion GFFs are obtained by inserting a complete set of intermediate states. 
In the region $t_\pi<t<16m_\pi^2$, only the $\pi\pi$ intermediate states contribute to the discontinuity (spectral function) of the GFFs, where we use the notation $t_i=4m_i^2$ with $i=\pi,K$ and $N$. 
In this case, the spectral function can be computed using the elastic unitarity condition via the Cutkosky cutting rule~\cite{Cutkosky:1960sp}, as illustrated in Fig.~\ref{fig.cut_pi}. 

The discontinuity reads
\begin{align}\label{eq.disc_piv1}
    &\operatorname{Disc}\left\langle\pi^a(p^\prime) \pi^b\left(p\right)\left|\hat{T}^{\mu \nu}(0)\right| 0\right\rangle \nonumber\\
    =\ &\frac{\delta^{ab}}{2}\left[\operatorname{Disc}A^\pi(t)\Delta^\mu \Delta^\nu +\operatorname{Disc}D^\pi(t)\left(P^\mu P^\nu-t g^{\mu \nu}\right) \right] \nonumber\\
    =\ & 2i\frac{2p_\pi}{\sqrt{t}}\frac{\delta^{ab}}{2}\bigg[\left(A^\pi(t)\right)^*\bigg(\frac{4}{3t}p_\pi^2(t^0_0(t)-t^0_2(t))\left(P^\mu P^\nu-tg^{\mu\nu}\right)+t^0_2(t)\Delta^\mu\Delta^\nu\bigg)\notag\\
    &\hspace{2cm}+\left(D^\pi(t)\right)^*t^0_0(t)\left(P^\mu P^\nu-t g^{\mu\nu}\right)\bigg] .
\end{align}
Here we use the standard $\pi \pi$ partial-wave amplitude notation $t_J^I$ for isospin $I$ and angular momentum $J$~\cite{Ananthanarayan:2000ht}. For a detailed derivation of Eq.~\eqref{eq.disc_piv1}, we refer to the Supplemental Material of Ref.~\cite{Cao:2024zlf}.

From Eq.~\eqref{eq.disc_piv1}, the spectral functions read
\begin{align}
    \operatorname{Im}A^\pi(t) & =\frac{2p_\pi}{\sqrt{t}}\left(t^0_2(t)\right)^* A^\pi(t)\ ,\label{eq.ImApi}\\
    \operatorname{Im}D^\pi(t) & =\frac{2p_\pi}{\sqrt{t}}\left[\frac{4}{3}\frac{p_\pi^2}{t}\left(t^0_0(t)-t^0_2(t)\right)^* A^\pi(t)+\left(t^0_0(t)\right)^* D^\pi(t)\right] .\label{eq.ImDpi}
\end{align}

It is noted from Eq.~\eqref{eq.ImApi} that $A^\pi$ carries the information of the well-defined $J^{PC}=2^{++}$ channel, while $D^\pi$ mixes $0^{++}$ and $2^{++}$ contributions according to Eq.~\eqref{eq.ImDpi}. 
Fortunately, the matrix elements of the symmetric rank-two tensor $\hat T^{\mu \nu}$ can be decomposed into a sum of two separately conserved irreducible tensors corresponding to well-defined $J^{P C}$, $0^{++}$ and $2^{++}$. 
Namely, the matrix element of EMT can be decomposed into a scalar trace part and a tensor traceless part as~\cite{Raman:1971jg}\footnote{The decomposition has also been used in Refs.~\cite{Broniowski:2024oyk,Broniowski:2025ctl}.}
\begin{align}
&\left\langle\pi^a(p^\prime)\pi^b\left(p\right)\left|\hat{T}^{\mu \nu}(0)\right| 0\right\rangle=\delta^{ab}\left(T_S^{\mu \nu}+T_T^{\mu \nu}\right) , \label{eq.EMT.ST}
\end{align}
where the scalar and tensor parts are~\cite{Cao:2024zlf}
\begin{align}
&T_S^{\mu \nu}=\frac{1}{3}\left(g^{\mu \nu}-\frac{P^\mu P^\nu}{P^2}\right) \Theta^\pi(t)\ , \\
&T_T^{\mu \nu}=T^{\mu \nu}-\frac{1}{3}\left(g^{\mu \nu}-\frac{P^\mu P^\nu}{P^2}\right) \Theta^\pi(t)=\left[\Delta^\mu \Delta^\nu+\frac{\Delta^2}{3t}\left(P^\mu P^\nu-tg^{\mu \nu}\right)\right] A^\pi(t)\ .\label{eq.Tensor_T}
\end{align}
Here, the trace FF $\Theta^\pi(t)$ is a pure $0^{++}$ (scalar) GFF and is defined as
\begin{align}
    \left\langle\pi^a(p^\prime) \pi^b\left(p\right)\left|\hat{T}^{\mu}_{\ \mu}(0)\right| 0\right\rangle &= \delta^{ab}\Theta^\pi(t) , 
\end{align}
Taking the trace of Eq.~\eqref{eq.pi_time} and comparing it with the above equation, we obtain the following relation
\begin{align}
    \Theta^\pi(t)=-\frac{1}{2}\left(4p_\pi^2 A^\pi(t)+3t D^\pi(t)\right) . \label{eq.Theta.FF}
\end{align}
The spectral function $\operatorname{Im}\Theta^\pi$ reads~\cite{Cao:2024zlf}
\begin{align}\label{eq.ImThetapiv1}
    \operatorname{Im}\Theta^\pi(t)=\frac{2p_\pi}{\sqrt{t}}\left(t^0_0(t)\right)^*\Theta^\pi(t)\ .
\end{align}

In order to take into account the final-state interactions in the $0^{++}$ channel between the $\pi\pi$ and $K\bar{K}$ states, the above unitarity relation needs to be generalized to a matrix form as~\cite{Donoghue:1990xh}
\begin{align}\label{eq.ImThetapiv2}
{\rm Im} \mathbf{\Theta}(t) = [\mathbf{T}^0_0(t)]^*\,\mathbf{\Sigma}_0^0(t) \,\mathbf{\Theta}(t)\ ,
\end{align}
where 
\begin{align}
    \mathbf{\Theta}(t) = \begin{pmatrix}
        \Theta^\pi(t) \\
        \frac{2}{\sqrt{3}}\Theta^K(t)
    \end{pmatrix} , \quad \mathbf{\Sigma}_0^0(t)=
    \begin{pmatrix}
        \sigma_\pi\theta(t-t_\pi) & 0 \\
        0 & \sigma_K\theta(t-t_K)
    \end{pmatrix} ,
\end{align}
with the kaon trace GFF $\Theta^K(t)=-\left[4p_K^2 A^K(t)+3t D^K(t)\right]/2$ and the phase-space factor $\sigma_i={2p_i}/{\sqrt{t}}, i\in\{\pi, K\}$. 
The couple-channel $\pi\pi$-$K\bar{K}$ scattering amplitudes in $IJ=00$ partial-wave are parametrized as
\begin{align}
  \mathbf{T}^0_0(t)= \left(\begin{array}{cc}
    \frac{\eta^0_0(t) e^{2 i \delta^0_0(t)}-1}{2 i \sigma_\pi} & |g^0_0(t)| e^{i \Psi^0_0(t)} \\
    |g^0_0(t)| e^{i \Psi^0_0(t)} & \frac{\eta^0_0(t) e^{2 i(\Psi^0_0(t)-\delta^0_0(t))}-1}{2 i \sigma_K}
\end{array}\right) , \label{eq.T00}
\end{align}
where the inelasticity parameter $\eta^0_0(t)$ can be related to the partial-wave $\pi\pi\rightarrow K\bar{K}$ amplitude $g^0_0(t)$ via $\eta^0_0(t)=\sqrt{1-4 \sigma_\pi \sigma_K|g^0_0(t)|^2 \theta\left(t-t_K\right)}$.

For the $2^{++}$ channel, analogously to Eq.~\eqref{eq.ImApi}, the single-channel unitarity relation of $A^K$ can be written as
\begin{align}\label{eq.ImAK}
\operatorname{Im}A^K(t)=\frac{\sqrt{3}p_\pi}{\sqrt{t}}p_\pi^4\left(g^0_2(t)\right)^* A^\pi(t)\ ,
\end{align}
where $g^0_2(t)$ is the $D$-wave isoscalar amplitude for $\pi\pi\rightarrow K\bar K$.
One can also generalize Eqs.~\eqref{eq.ImApi} and~\eqref{eq.ImAK} to the coupled-channel case as
\begin{align}\label{eq.ImApiv2}
 \operatorname{Im} \mathbf{A}(t)=   
   [\mathbf{T}^0_2(t)]^*\,\mathbf{\Sigma}_2^0(t) \,\mathbf{A}(t)\ ,
\end{align}
where the $2^{++}$ GFFs and phase space factors are
\begin{align}
 \mathbf{A}(t) =\left(\begin{array}{c}
    A^\pi(t) \\
    \frac{2}{\sqrt{3}} A^K(t)
    \end{array}\right)  \ ,\quad \mathbf{\Sigma}_2^0(t)=
\left(\begin{array}{cc}
    \sigma_\pi p_\pi^4\theta(t-t_\pi) & 0 \\
    0 & \sigma_K p_K^4\theta(t-t_K)
    \end{array}\right) .\label{eq.sigma20}
\end{align}
Similarly to Eq.~\eqref{eq.T00}, the $IJ=02$ $\pi\pi$-$K\bar{K}$ coupled-channel scattering amplitudes are parametrized as
\begin{align}
\mathbf{T}^0_2=\left(\begin{array}{cc}
    \frac{\eta^0_2(t) e^{2 i \delta^0_2(t)}-1}{2 i \sigma_\pi p_\pi^4} & |g^0_2(t)| e^{i \Psi^0_2(t)} \\
    |g^0_2(t)| e^{i \Psi^0_2(t)} & \frac{\eta^0_2(t) e^{2 i(\Psi^0_2(t)-\delta^0_2(t))}-1}{2 i \sigma_K p_K^4}
\end{array}\right) ,\label{eq.T20}
\end{align}
with $\eta^0_2(t)=\sqrt{1-4 \sigma_\pi \sigma_K (p_\pi p_K)^4|g^0_2(t)|^2 \theta\left(t-t_K\right)}$.

\subsection{Coupled-channel Muskhelishvili-Omn\` es formalism}

In Ref.~\cite{Cao:2024zlf}, the GFF $A^\pi$ in the single-channel case, i.e., Eq.~\eqref{eq.ImApi}, has already been thoroughly analyzed using the Omn\`es formalism. 
Building upon the methodology developed for the trace GFF analysis in~\cite{Cao:2024zlf}, we extend the analysis to the $D$-wave coupled-channel case. 

First, we need to provide the $D$-wave Omn\`es matrix in the coupled-channel scenario, i.e., Eq.~\eqref{eq.ImApiv2}.
The $D$-wave Omn\`es matrix corresponds to the so-called Muskhelishvili-Omn\` es (MO) solution~\cite{Muskhelishvili:1953,Omnes:1958hv} to the following homogeneous integral equation: 
\begin{align}\label{eq.MO-D}
    \boldsymbol{\Omega}_{2}^0(t)
    = \frac{1}{\pi}\int^\infty_{t_\pi}\frac{\md t^\prime}{t^\prime-t} [\mathbf{T}_2^0(t)]^* \,\mathbf{\Sigma}_2^0(t)\,\boldsymbol{\Omega}_{2}^0(t^\prime)\ ,
\end{align}
with $\mathbf{T}_2^0(t)$ and $\mathbf{\Sigma}_2^0(t)$ given by Eqs.~\eqref{eq.sigma20} and~\eqref{eq.T20}, respectively. 
As for the inputs, the $D$-wave phase shift is taken from the latest crossing-symmetric dispersive analysis~\cite{Bydzovsky:2016vdx}, and the $D$-wave amplitude modulus $|g_2^0|$ and phase $\Psi^0_2$ are taken from the results of the CFD parameter set in Ref.~\cite{Pelaez:2020gnd}.  
We extrapolate the phase shift beyond $E_0\simeq 2~$GeV via the a smooth function connecting the value at $E_0$ to the assumed asymptotic value $\pi$ in the following form~\cite{Moussallam:1999aq}:
\begin{align}\label{eq.delta_extra}
\delta_2^0(t)=\pi+\left(\delta_2^0\left(E_0^2\right)-\pi\right) f_\delta\!\left(\frac{\sqrt{t}}{E_0}\right) ,
\end{align}
with $f_\delta(x)=2/(1+x^3)$. 
The same extrapolation strategy is also applied to $|g_2^0|$ and $\Psi^0_2$, but with asymptotic values $|g_2^0(\infty)|=0$ and $\Psi^0_2(\infty)=2\pi$. 

\begin{figure}[t]
    \centering\includegraphics[width=1\textwidth,angle=-0]{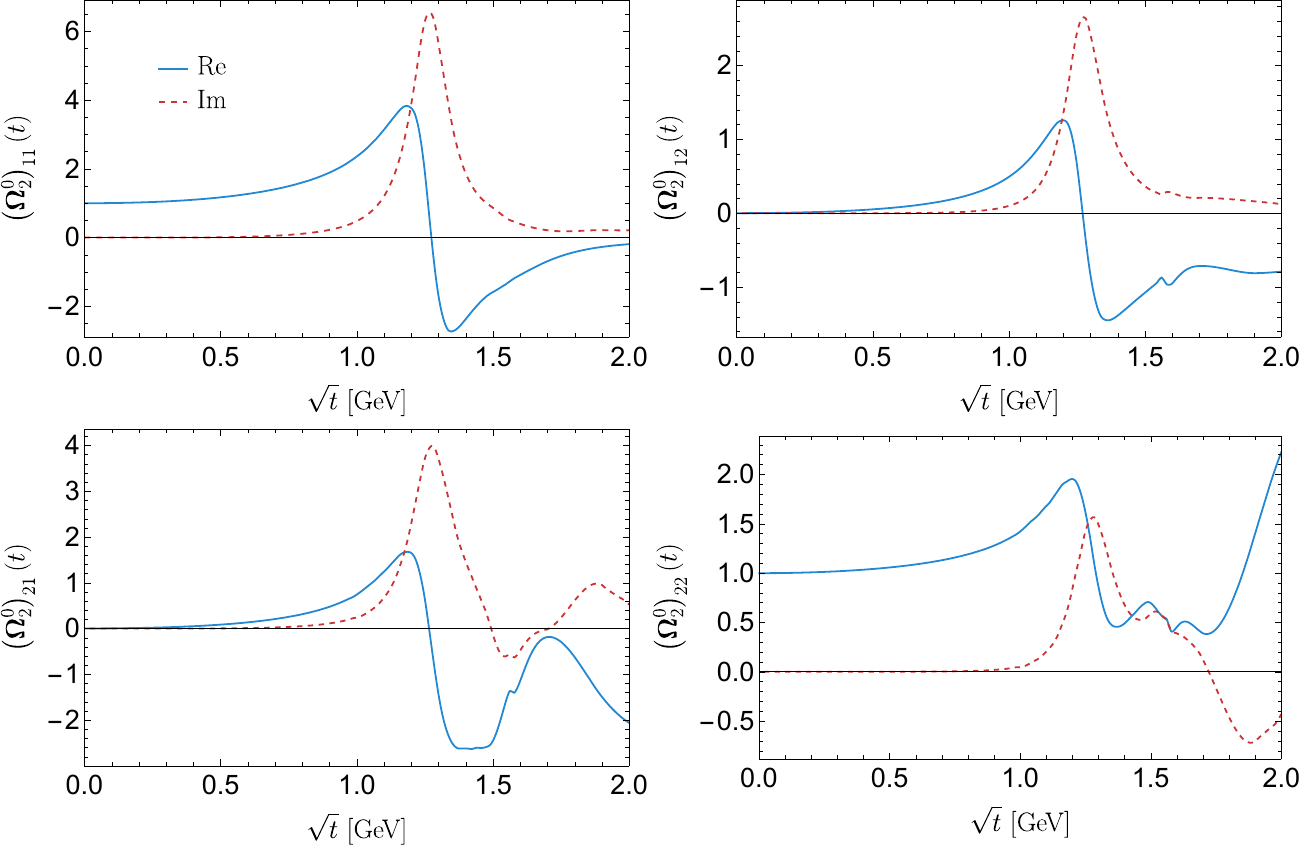} 
    \caption{Results for real and imaginary parts of the components of the $D$-wave Omn\`es matrix.} \label{fig.MO02Matrix}
\end{figure}

\begin{figure}[t]
    \centering\includegraphics[width=.5\textwidth,angle=-0]{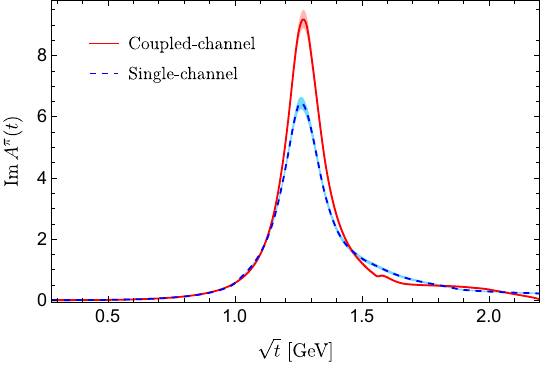}   
    \caption{Results of the spectral function $\operatorname{Im}A^\pi(t)$.} \label{fig.ImApi}
\end{figure}

The solution of Eq.~\eqref{eq.MO-D} is obtained numerically using the discretization procedure described in Ref.~\cite{Moussallam:1999aq}. 
The $D$-wave Omn\` es matrix is shown in Fig.~\ref{fig.MO02Matrix}.\footnote{Note that our results differ from the $D$-wave Omn\` es matrix in Ref.~\cite{TarrusCastella:2021pld}, primarily due to different choices for the asymptotic behavior of the $D$-wave phase shift. 
$\delta_2^0\rightarrow 2\pi$ was used in Ref.~\cite{TarrusCastella:2021pld}, while we choose $\delta_2^0\rightarrow \pi$ as shown in Eq.~\eqref{eq.delta_extra}.
We checked the impact of choosing different asymptotic behaviors of the phase shift on GFFs $A^{\pi,K}$, and found that the difference is smaller than the corresponding uncertainties at the current level of precision. }
The GFFs are related to the Omn\`es matrix by
\begin{align}
   \left[\mathbf{A}(t)\right]^{\rm T} = \left[\mathbf{P}_2(t)\right]^{\rm T} \mathbf{\Omega}_2^0(t)\ , \quad \mathbf{P}_2(t) =\left(\begin{array}{c}
    1+\alpha_\pi t \\
    \frac{2}{\sqrt{3}}\left(1+\alpha_K t\right)
    \end{array}\right) ,
\end{align}
with $\mathbf{A}(t)$ defined in Eq.~\eqref{eq.sigma20} and $\alpha_{\pi, K}$ two parameters. More explicitly, one has 
\begin{align}\label{eq.Api_couple}
\begin{aligned}
& A^\pi(t)=\left(1+\alpha_\pi t\right) \left(\boldsymbol{\Omega}_{2}^0\right)_{11}(t)+\frac{2}{\sqrt{3}}\left(1+\alpha_K t\right) \left(\boldsymbol{\Omega}_{2}^0\right)_{12}(t)\ , \\
& A^{K}(t)=\frac{\sqrt{3}}{2}\left(1+\alpha_\pi t\right) \left(\boldsymbol{\Omega}_{2}^0\right)_{21}(t)+\left(1+\alpha_K t\right) \left(\boldsymbol{\Omega}_{2}^0\right)_{22}(t)\ ,
\end{aligned}
\end{align}
where the parameters $\alpha_{\pi, K}$ are related to the slopes at $t=0$,
\begin{align}
\begin{aligned}
& \alpha_\pi=\dot{A}^\pi(0)- \left(\dot{\boldsymbol{\Omega}}_{2}^0\right)_{11}(0)-\frac{2}{\sqrt{3}} \left(\dot{\boldsymbol{\Omega}}_{2}^0\right)_{12}(0)\ , \\
& \alpha_K=\dot{A}^{K}(0)-\frac{\sqrt{3}}{2} \left(\dot{\boldsymbol{\Omega}}_{2}^0\right)_{21}(0)-\left(\dot{\boldsymbol{\Omega}}_{2}^0\right)_{22}(0)\ ,
\end{aligned}
\end{align}
where we have used $\left(\boldsymbol{\Omega}_2^0\right)_{ij}(0)=\delta_{ij}$, and $\dot{A}^{\pi,K}(0)=-2L_{12}^r/F_\pi^2$ according to the $\mathcal{O}(p^4)$, i.e., NLO ChPT result of $A(t)$~\cite{Donoghue:1991qv}
\begin{align}\label{eq.Aslope}
    A^i(t)=1-\frac{2L^r_{12}}{F_\pi^2}t+\mathcal{O}(p^4)\ ,\quad i=\pi,K,\eta\ .
\end{align}
The LEC $L^r_{12}$ can be estimated by resonance saturation. 
The tensor meson dominance (TMD) model gives~\cite{Donoghue:1991qv}
\begin{align}\label{eq.TMD}
    A^\pi(t)=\frac{m_{f_2}^2}{m_{f_2}^2-t}=1+\frac{t}{m_{f_2}^2}+\cdots
\end{align}
where $m_{f_2}$ is the mass of the lowest-lying tensor meson $f_2(1270)$, which leads to $L_{12}^r=-F_\pi^2/(2m_{f_2}^2)$. 
It should be noted that slightly different values have been reported for the $f_2(1270)$ mass, namely $(1259 \pm 4 \pm 4)~$MeV~\cite{PhysRevD.91.052006}, $(1263 \pm 12)~$MeV~\cite{CLAS:2020ngl}, $(1275 \pm 6)~$MeV~\cite{Klempt:2022qjf}, and the averaged value in the Review of Particle Physics, $m_{f_2}=(1275.4\pm 0.8)~$MeV~\cite{ParticleDataGroup:2024cfk}.
Here, we set $m_{f_2}=(1275 \pm 20)~$MeV to cover all these values for error estimation.
For illustration, we show the spectral function $\operatorname{Im}A^\pi$ in Fig.~\ref{fig.ImApi} for the coupled-channel case \eqref{eq.Api_couple}, with the single-channel analogue from Ref.~\cite{Cao:2024zlf} also shown for comparison.

\begin{figure}[t]
    \centering\includegraphics[width=1\textwidth,angle=-0]{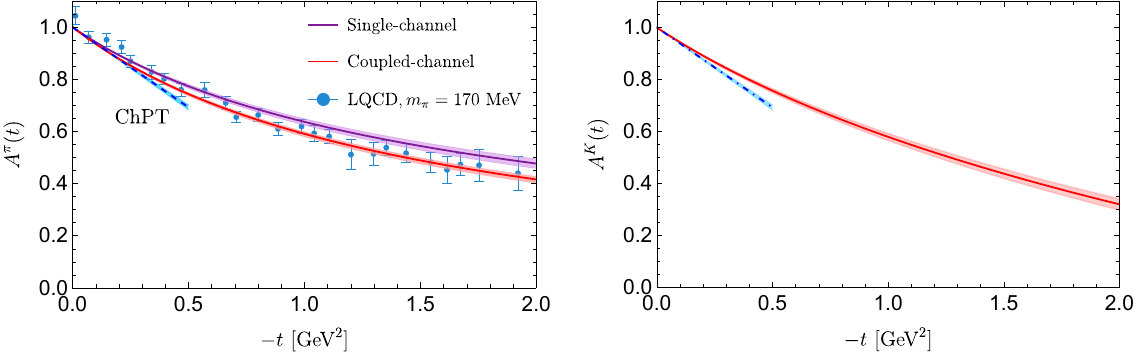}   
    \caption{The total GFFs $A^{\pi/K}$ from single-channel Omn\`es solution~\cite{Cao:2024zlf} and coupled-channel MO solution. The blue dashed lines show the NLO ChPT prediction in the small-$|t|$ region. We also show the LQCD results at $m_\pi=170~$MeV in Ref.~\cite{Hackett:2023nkr} for comparison.} \label{fig.Api,K}
\end{figure}
The coupled-channel results for the GFFs $A^{\pi/K}$ are also shown in Fig.~\ref{fig.Api,K}. 
Note that our results agree well with the LQCD calculations at a pion mass of 170~MeV~\cite{Hackett:2023nkr}, which is slightly larger than the physical value.
However, a discrepancy exists between the single-channel and coupled-channel GFF $A^\pi$. 
This discrepancy implies that the high-energy parts of the $D$-waves for $\pi\pi \rightarrow \pi\pi$~\cite{Bydzovsky:2016vdx} and $\pi\pi \rightarrow K\bar{K}$~\cite{Pelaez:2020gnd} play a visible role in the $D$-wave GFF $A^\pi$. 
Refined phase shift analyses are necessary to further improve these results in the future.

\begin{figure}[t]
    \centering\includegraphics[width=.5\textwidth,angle=-0]{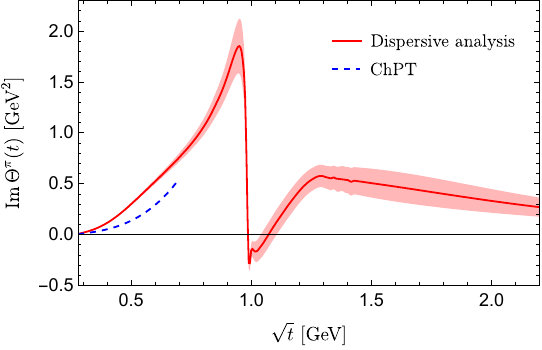}   
    \caption{Results of the spectral function $\operatorname{Im}\Theta^\pi(t)$. The blue dashed line shows the NLO ChPT prediction in small-$|t|$ region.} \label{fig.ImThetapi}
\end{figure}

The same procedure has been applied to the $S$-wave GFFs $\Theta^{\pi,K}$~\cite{Cao:2024zlf}, for which a coupled-channel analysis is expected to be reliable at low energies. 
To be self-contained, we also provide the coupled-channel framework for the pion and kaon trace GFFs that has been developed in Ref.~\cite{Cao:2024zlf}.
The scalar GFFs can be obtained through~\cite{Donoghue:1990xh}\footnote{It is important to note that the polynomial $\boldsymbol{P}_0(t)$ cannot be constant due to the chiral symmetry constraints as discussed in Ref.~\cite{Donoghue:1990xh}.}
\begin{align}
 [\boldsymbol{\Theta}(t)]^T=[\boldsymbol{P}_0(t)]^T \boldsymbol{\Omega}_0^0(t)\ ,\quad    \boldsymbol{P}_0(t) =\left(\begin{array}{c}
    2 m_\pi^2+\beta_\pi t \\
   \frac{2}{\sqrt{3}}\left(2 m_K^2 +\beta_K t\right)
    \end{array}\right) .
\end{align}
Using $\left(\boldsymbol{\Omega}_0^0\right)_{ij}(0)=\delta_{ij}$, one gets the parameters $\beta_\pi, \beta_K$ as:
\begin{align}
\begin{aligned}
& \beta_\pi=\dot{\Theta}^\pi(0)-2 m_\pi^2 \left(\dot{\boldsymbol{\Omega}}_{0}^0\right)_{11}(0)-\frac{4 m_{K}^2}{\sqrt{3}} \left(\dot{\boldsymbol{\Omega}}_{0}^0\right)_{12}(0)\ , \\
& \beta_K=\dot{\Theta}^{K}(0)-\sqrt{3} m_\pi^2 \left(\dot{\boldsymbol{\Omega}}_{0}^0\right)_{21}(0)-2 m_K^2 \left(\dot{\boldsymbol{\Omega}}_{0}^0\right)_{22}(0)\ ,
\end{aligned}
\end{align}
and the slopes $\dot{\Theta}^{\pi,K}(0)$ can be obtained in ChPT at NLO~\cite{Donoghue:1991qv}: 
\begin{align}
    \dot{\Theta}^\pi(0)&=1-4L_{12}^r\frac{m_\pi^2}{F_\pi^2}-24 L^r \frac{m_\pi^2}{F_\pi^2}-\frac{3}{2}\frac{m_\pi^2}{F_\pi^2} I_\pi+\frac{m_\pi^2}{2 F_\pi^2} I_\eta=0.98(2)\ ,\label{eq.slope_thetapi}\\ 
    \dot{\Theta}^K(0)&=1-4L_{12}^r\frac{m_K^2}{F_\pi^2}-24 L^r \frac{m_K^2}{F_\pi^2}-\frac{m_K^2}{F_\pi^2} I_\eta=0.94(14)\ .
\end{align}
The spectral function $\operatorname{Im}\Theta^\pi(t)$ is shown in Fig.~\ref{fig.ImThetapi}. In the vicinity of the $K\bar K$ threshold, where the scalar $f_0(980)$ resonance is located, $\operatorname{Im} \Theta^\pi(t)$ drops sharply below zero, consistent with previous analyses~\cite{Donoghue:1990xh, Broniowski:2024oyk}. 
One sees that from 1.2 to 2.2~GeV, $\operatorname{Im} \Theta^\pi(t)$ does not exhibit the alternating positive and negative pattern observed in Ref.~\cite{Broniowski:2024oyk}. 
This is primarily because we employ a simple extrapolation similar to Eq.~\eqref{eq.delta_extra} for the $S$-wave phase shift above 1.3~GeV, where the $\pi \pi, K \bar{K}$ coupled-channel approximation becomes invalid.
For a detailed analysis of the high-energy behavior of the spectral function of the GFF $\Theta^\pi$, see Refs.~\cite{Broniowski:2024oyk,Broniowski:2024mpw}.

\begin{figure}[t]
    \centering\includegraphics[width=1\textwidth,angle=-0]{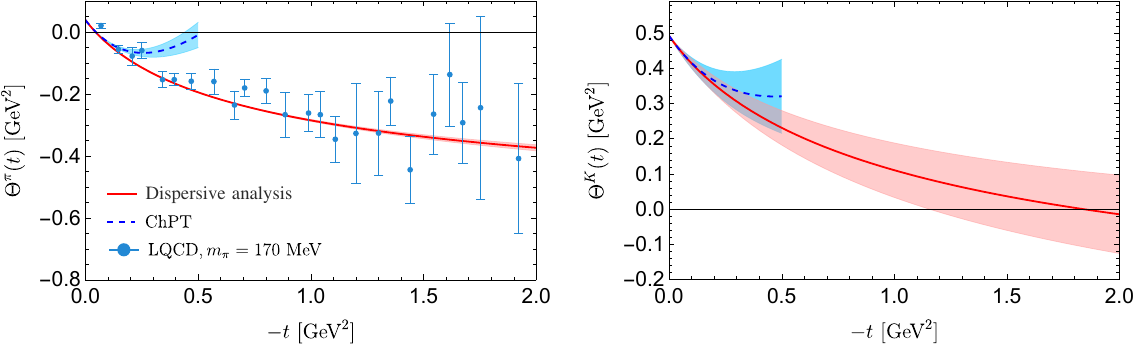}
    \caption{The total GFFs $A^{\pi/K}$ of the pion and kaon from Ref.~\cite{Cao:2024zlf}. The blue dashed lines show the NLO ChPT prediction in small-$|t|$ region. The LQCD results for $\Theta^\pi$ at $m_\pi=170~$MeV are obtained from a linear combination of  $A^\pi$ and $D^\pi$ in Ref.~\cite{Hackett:2023nkr}, with errors added in quadrature.} \label{fig.ThetapiK}
\end{figure}

The corresponding results for the trace pion and kaon GFFs~\cite{Cao:2024zlf} are shown in
Fig.~\ref{fig.ThetapiK}.

\begin{figure}[ht]
    \centering\includegraphics[width=1\textwidth,angle=-0]{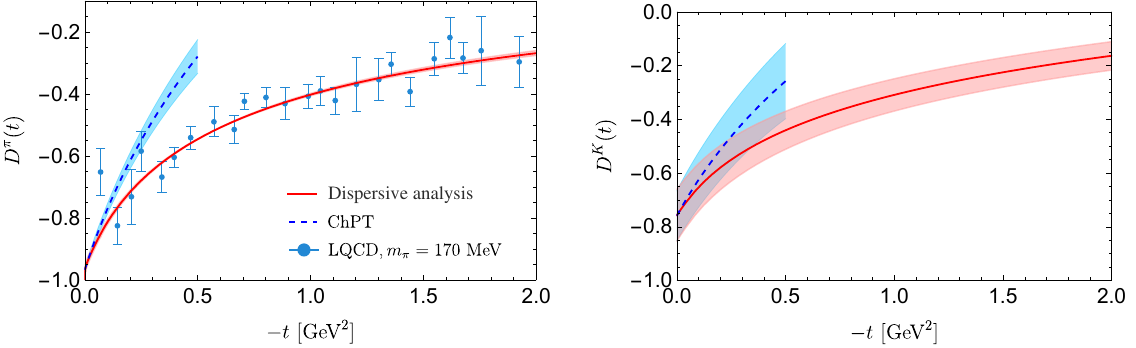} 
    \caption{The total GFFs $D^{\pi/K}$ of the pion and kaon. The blue dashed lines show the NLO ChPT prediction in the small-$|t|$ region. We also show the LQCD data at $m_\pi=170~$MeV in Ref.~\cite{Hackett:2023nkr} for comparison.} \label{fig.DpiK}
\end{figure}

With the phase shift inputs~\cite{Ananthanarayan:2000ht,Bydzovsky:2016vdx}, the GFFs $D^{\pi,K}(t)=-2\left({\Theta^{\pi,K}(t)+2p^2_{\pi,K}A^{\pi,K}(t)}\right)/(3t)$ can be obtained as displayed in Fig.~\ref{fig.DpiK}.
We have checked that the statistical errors from phase shifts caused by the $D$-wave and $S$-wave Omn\`es matrices are negligible.\footnote{In practice, the numerical Omn\`es matrices, or the phase shifts used by different groups, are not identical. 
For example, the Bern group~\cite{Hoferichter:2012wf} and the Madrid-Krakow group~\cite{Garcia-Martin:2011iqs} employ different $S$-wave phase shifts, with the former being implemented in the present work. 
These phase shifts differ in the energy region near $1$~GeV, leading to sizable difference in Omn\`es matrix elements such as $\left(\boldsymbol{\Omega}_0^0\right)_{21}$.
The uncertainties arising from using different $S$-wave phase shifts are of the same order of magnitude as the statistical errors inherent in the phase shifts themselves. Therefore, we do not account for the errors caused by this issue, as this effect is nearly negligible in our context. A more detailed discussion can be found in Ref.~\cite{Blackstone:2024ouf}.} 
Our results for the pion GFFs show good agreement with recent LQCD simulations at $m_\pi=170$~MeV~\cite{Hackett:2023nkr}. 
The predicted kaon GFFs can be checked with future lattice QCD calculations.

\subsection{Pion mass dependence of the pion GFFs}

The pion mass dependence has been studied in the literature for the pion vector FF~\cite{Guo:2008nc}, which is connected to the corresponding $P$-wave $\pi\pi$ scattering amplitudes using the Omn\`es representation. 
Here, we aim to present the pion mass dependence of the pion GFFs. 
For the trace GFF $\Theta^\pi$, we rely on the prediction of $\dot{\Theta}^{\pi}(0)$ from $\mathcal{O}(p^4)$ ChPT and the $S$-wave phase shifts at unphysical pion masses from crossing-symmetric Roy-equation analyses~\cite{Cao:2023ntr, Rodas:2023nec}.
The Roy-equation formalism is model-independent and does not depend on the unitarization method as used in Ref.~\cite{Guo:2008nc} for the pion vector FF.
However, for the tensor GFF $A^\pi$, given the lack of precise $D$-wave phase shifts at unphysical pion masses, we adopt the TMD model in Eq.~\eqref{eq.TMD} instead. 
We also notice that the pion mass effects have been recently discussed in the pion GFF case~\cite{Broniowski:2024oyk,RuizArriola:2024udm} using ChPT with the $\overline{\rm MS}$ scheme to compare with the lattice QCD results at $m_\pi=170$~MeV~\cite{Hackett:2023nkr}.

We consider three unphysical pion masses, namely $m_\pi = 239, 283, 391$~MeV, for which dispersive analyses exist.
Note that when $m_\pi \gtrsim 300$~MeV, the $f_0(500)$ (also known as $\sigma$) pole becomes a bound state pole on the first Riemann sheet of the complex energy plane~\cite{Cao:2023ntr, Rodas:2023nec}. 
In the single-channel scenario, the dispersive representation of $\Theta^\pi(t)$ can be written as
\begin{align}
    \Theta^\pi(t)=P_0^\pi(t) \Omega_{0}^{0}(t), \quad \Omega_{0}^{0}(t) \equiv \exp \left\{\frac{t}{\pi} \int_{t_\pi}^{\infty} \frac{\mathrm{d} t^{\prime}}{t^{\prime}} \frac{\delta_0^0\left(t^{\prime}\right)}{t^{\prime}-t}\right\} ,
\end{align}
where the $S$-wave phase shift at the physical pion mass can be taken from Ref.~\cite{Garcia-Martin:2011iqs} and that at unphysical pion masses can be taken from Refs.~\cite{Cao:2023ntr, Rodas:2023nec}. 

For $m_\pi=239$ and $283~$MeV, we choose the matching point between the input phase shift and the extrapolation at $E_0\simeq 0.9$~GeV, which is below the $f_0(980)$ and the $K\bar{K}$ threshold. Similarly to Eq.~\eqref{eq.delta_extra}, the extrapolation to the asymptotic value $\pi$ at $t\rightarrow +\infty$ is done via
\begin{align}\delta_0^0(t)=\pi+\left(\delta_0^0\left(E_0^2\right)-\pi\right) f\!\left(\frac{\sqrt{t}}{E_0}\right) .
\end{align}

\begin{figure}[ht]
    \centering\includegraphics[width=1.0\textwidth,angle=-0]{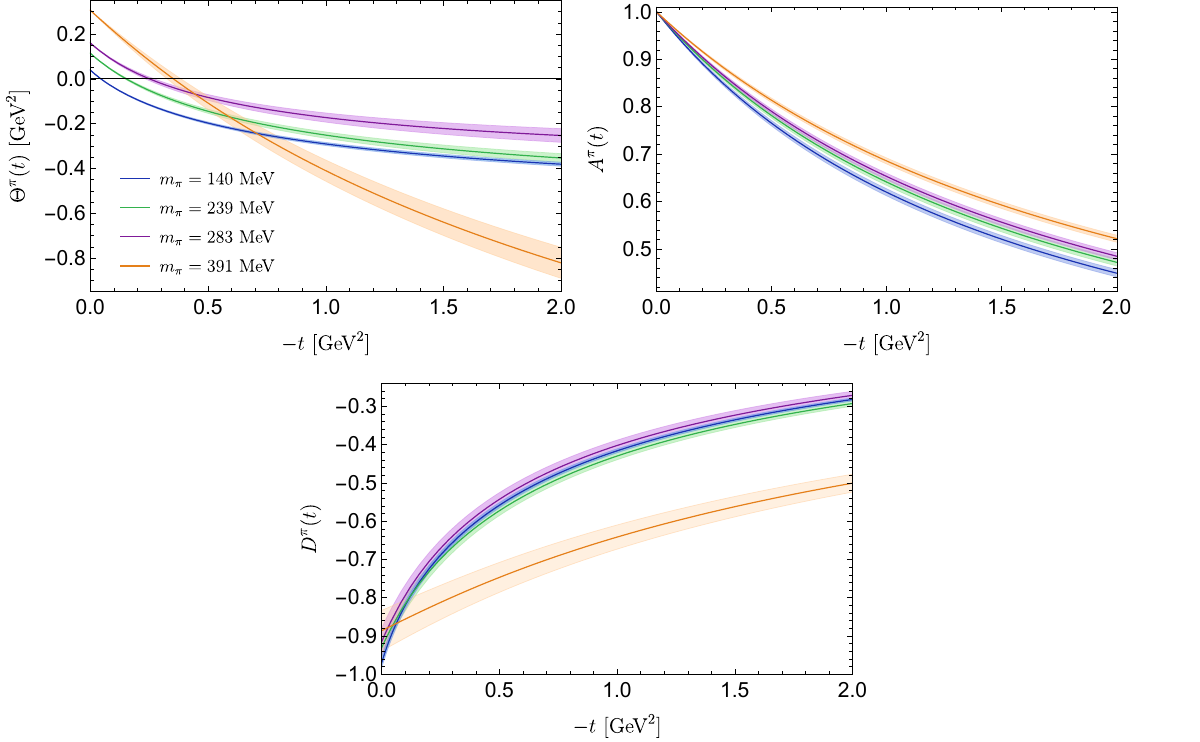}
    \caption{Pion mass dependence of the pion GFFs.} \label{fig.mpi_dependence}
\end{figure}

For $m_\pi = 391$~MeV, the distance between the $\pi\pi$ and $K\bar{K}$ thresholds is small, where the corresponding kaon mass is $549$~MeV, and the extrapolation of the phase shifts is handled slightly differently. 
We also adopt the single-channel framework where $E_0\simeq 1.4$~GeV and take into account the inelastic contributions of the $K\bar{K}$ channel by replacing the $S$-wave phase shift to the phase of the $\pi\pi$ $S$-wave scattering amplitude, $\phi^0_0(t)$.
The polynomial $P^\pi_0$ is set to
\begin{align}P^\pi_0(t)=2m_\pi^2+\left(\dot{\Theta}^\pi(0)-\frac{2m_\pi^2}{\pi} \int_{t_\pi}^{\infty} \mathrm{d} t^{\prime} \frac{\delta_0^0\left(t^{\prime}\right)}{t^{\prime 2}}\right)t\ .
\end{align}
Equation~\eqref{eq.slope_thetapi} gives the slope $\dot{\Theta}^\pi(0)$ at $t=0$, whose $m_\pi$ dependence is determined by the $m_\pi$ dependence of $m_K,m_\eta$ and $F_\pi$~\cite{Nebreda:2010wv}. 
We take the values of the LECs fitted at $\mathcal{O}(p^4)$, which are given in Table~6 of Ref.~\cite{Bijnens:2014lea}. The results of $\dot{\Theta}^\pi(0)$ at various pion masses are listed in Table~\ref{tab.slope}.
\begin{table}[tb]
    \centering
    \caption{The $\mathcal{O}(p^4)$ ChPT predictions of the slope $\dot{\Theta}^\pi(0)$ at various pion masses.}
    \begin{tabular}{l | cccc}
    \hline\hline
        $m_\pi$ & 140~MeV & 239~MeV & 283~MeV & 391~MeV \\ \hline
        $\dot{\Theta}^\pi(0)$ & $0.98(2)$ & $0.96(3)$ & $0.96(4)$ & $0.97(7)$ \\
    \hline\hline
    \end{tabular}
    \label{tab.slope}
\end{table}
Interestingly, the dependence of $\dot{\Theta}^\pi(0)$ on $m_\pi$ is not monotonic. 
Figure~\ref{fig.mpi_dependence} shows the results of the spacelike GFF $\Theta^\pi(t)$ with varying $m_\pi$. 

Some comments are in order. 
The coupled-channel GFF $\Theta^\pi$ in Fig.~\ref{fig.ThetapiK} and the single-channel one agree with each other within errors.
This confirms the validity of the use of the single-channel formalism here. 
One can also clearly observe that for the pion mass between 283~MeV and $391$~MeV, the GFF $\Theta^\pi(t)$ undergoes notable changes. This should be owing to the fact that the $f_0(500)$ state becomes a bound state pole at a pion mass between those two values.
At the pion mass where the $f_0(500)$ turns into an $S$-wave $\pi\pi$ bound state exactly at the threshold, the trace GFF $\Theta^\pi(t)$ at a given $t$ must exhibit a threshold cusp in the pion mass dependence. 
This cusp causes the $\Theta^\pi$ value at larger pion masses to change more rapidly than those at lower pion masses.
Similar singularities in $m_\pi$ dependence have been discussed in Refs.~\cite{Guo:2011gc} for the pion vector FF and in Ref.~\cite{Guo:2012tg} in heavy quarknium physics.

The TMD model adequately describes the GFF $A^\pi$~\cite{Masjuan:2012sk, Broniowski:2024oyk} (cf. Eq.~\eqref{eq.TMD}), given that the $I=0$ $D$-wave $\pi\pi$ scattering is primarily governed by the $f_2(1270)$ resonance. 
Here, we use the latest prediction for the pion mass dependence of $m_{f_2}$ from the resonance chiral theory approach~\cite{Chen:2023ybr}. 
Given the small error associated with this prediction, we add a fixed uncertainty of $\pm 20$~MeV to $m_{f_2}$, as in the physical case. 
The calculated results are presented in Fig.~\ref{fig.mpi_dependence}. 
The variation of $A^\pi$ is quite mild as $m_\pi$ increases. 
By applying Eq.~\eqref{eq.Theta.FF}, we can also determine the behavior of $D^\pi$ as $m_\pi$ varies. 
Figure~\ref{fig.mpi_dependence} shows that $D^\pi$ undergoes a significant change when $m_\pi$ increases from 283 to 391~MeV, which is inherited from that of $\Theta^\pi(t)$. 
Thus, we can conclude that the mechanical properties of the pion exhibit minor changes for small pion masses. 
However, significant changes emerge when $m_\pi$ is large enough for the $\sigma$ meson to become a bound state.

\section{Nucleon gravitational form factors}

\subsection{Dispersive representation of nucleon gravitational form factors}
\label{sec.4}

Applications of the DR techniques to the nucleon GFFs were pioneered by Ref.~\cite{Pasquini:2014vua}, which established the first $t$-channel dispersive analysis of the quark $D$-term GFF of the nucleon in deeply virtual Compton scattering. 
The study in this pioneering work is limited by model-dependent estimates of the $\pi\pi$ generalized distribution amplitudes, neglect of $K\bar{K}$ intermediate states, and absence of systematic uncertainty quantification. 
These limitations have been overcome in Ref.~\cite{Cao:2024zlf} by including $S$-wave $\pi\pi$-$K\bar{K}$ coupled channels, utilizing partial waves from modern $\pi N$ Roy-Steiner equation analyses instead of the old Karlsruhe-Helsinki results~\cite{Hohler:1983}, and implementing systematic uncertainty estimates. 
This dispersive framework provides the first determination of complete nucleon GFFs with a reasonable error estimate.
In this subsection, we review the dispersive representation of the nucleon GFFs derived in Ref.~\cite{Cao:2024zlf}, setting up the basis for the results of the nucleon spatial density profiles to be presented in the next section.

The imaginary parts of the nucleon GFFs can be obtained by inserting a complete set of intermediate states via
\begin{align}\operatorname{Disc}\left\langle N(p^\prime) \bar N\left(p\right)\left|\hat{T}^{\mu \nu}(0)\right| 0\right\rangle\propto \sum_n\left\langle\left. N(p^\prime) \bar N\left(p\right)\right|n\right\rangle\left\langle n\left|\hat T^{\mu\nu}(0)\right|0\right\rangle^*\delta^4(p+p^\prime-p_n)\ ,
\end{align}
where $|n\rangle$ denotes asymptotic states with momentum $p_n$. 
The intermediate states should carry the same quantum numbers as the current $\hat T^{\mu\nu}$: $I^G\left(J^{P C}\right)=0^{+}\left(0^{++},2^{++}\right)$ for the isoscalar component, and $1^{-}\left(1^{++}\right)$ for the isovector component.
In the exact isospin limit, i.e., $m_u=m_d\equiv \hat m$, considered in Ref.~\cite{Cao:2024zlf} and here, the quark mass part of QCD EMT is given by 
\begin{align}
    \hat{T}^{\mu\nu} \supset g^{\mu\nu}\left(m_u \bar uu+m_d \bar dd+\cdots\right)=g^{\mu\nu}\hat{m}\left(\bar uu+\bar dd+\cdots\right) ,
\end{align}
which is a standard isoscalar current, and the isovector part proportional to $\left(m_u-m_d\right)$ has been neglected. 
In this case, the nucleon EMT matrix elements can only couple to the asymptotic states with quantum numbers $0^{+}\left(0^{++},2^{++}\right)$ ($2\pi, 4\pi, K \bar K, \cdots$).
Then, Eq.~\eqref{eq.isospinv1} is reduced to
\begin{align}
    X^p=X^n=X^s=X^N\ .
\end{align}
This corresponds to the physics picture that classical gravitational interactions do not distinguish between particles within the isospin multiplet, resulting in identical GFFs for the proton and neutron in the isospin limit.\footnote{For a discussion of the long-range electromagnetic effects of the $D$-term GFF, see Ref.~\cite{Varma:2020crx}.} 

\begin{figure}[t]
    \centering
    \includegraphics[width=.68\textwidth,angle=-0]{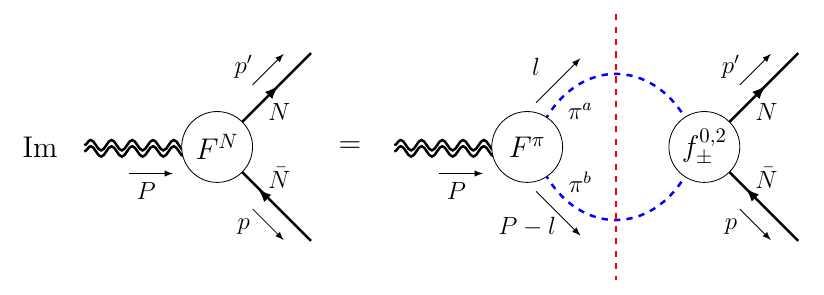}
    \caption{Elastic unitarity relation for the isoscalar nucleon GFFs $F^N=\{A^s,J^s,D^s\}$~\cite{Cao:2024zlf}. The blue dashed lines denote pions, the black solid lines stand for nucleons, the double wiggly lines represent the external QCD EMT current, and the dashed red vertical line indicates that the intermediate pions are on-shell.}\label{fig.cut_nucleon}
\end{figure}
In the region $t_\pi<t<16m_\pi^2$, only the $\pi\pi$ intermediate state contributes to the discontinuity of the nucleon GFF. In this situation, the spectral function can be derived using the Cutkosky cutting rule, as illustrated in Fig.~\ref{fig.cut_nucleon}, and the results are given by (for more details, see Ref.~\cite{Cao:2024zlf}),
\begin{align}
    \operatorname{Im}A^s(t) & =\frac{3p_\pi^5}{\sqrt{6t}}\left[f^2_-(t)+\sqrt{\frac{3}{2}}\frac{m_N}{p_N^2}\Gamma^2(t)\right]^* A^\pi(t)\ ,\label{eq.ImA_nucleon}\\
    \operatorname{Im}J^s(t) & =\frac{3p_\pi^5}{2\sqrt{6t}}\left(f^2_-(t)\right)^* A^\pi(t)\ ,\label{eq.ImJ_nucleon}\\
    \operatorname{Im}D^s(t) & =-\frac{3 m_N p_\pi}{2 p_N^2\sqrt{t}}\left[\frac{4p_\pi^2}{3t}\left(\left(f^0_+(t)\right)^*-\left(p_\pi p_N\right)^2\left(f^2_+(t)\right)^*\right)A^\pi(t)+\left(f^0_+(t)\right)^* D^\pi(t)\right] ,\label{eq.ImD_nucleon}
\end{align}
where $\Gamma^2(t)=m_N\sqrt{\frac{2}{3}} f^2_-(t)-f^2_+(t)$, $f_{\pm}^J(t)$ are the partial wave isospin-even (odd) $\pi\pi\rightarrow N\bar{N}$ scattering amplitudes~\cite{Frazer:1960zza,Frazer:1960zzb,Hohler:1983} (for Lorentz and isospin decomposition of the $\pi N$ scattering amplitudes, see, e.g., Ref.~\cite{Yao:2016vbz}).

Analogously to the pion GFFs, the nucleon matrix elements of $\hat T^{\mu \nu}$ can also be decomposed into separately conserved traceful scalar part ($J^{P C}=0^{++}$) and a traceless irreducible tensor part ($J^{P C}=2^{++}$). Namely,
\begin{align}
\left\langle N(p^\prime) \bar N(p)\left|\hat T^{\mu\nu}(0)\right|0\right\rangle=\bar u(p^\prime)\left(T_S^{\mu \nu}+T_T^{\mu \nu}\right)v(p)\ , 
\end{align}
with the trace and traceless parts given by~\cite{Cao:2024zlf, Broniowski:2025ctl}
\begin{align}
T_S^{\mu \nu}=&\frac{1}{3}\left(g^{\mu \nu}-\frac{P^\mu P^\nu}{P^2}\right) \Theta^s(t)\ , \\
T_T^{\mu \nu}=&T^{\mu \nu}-\frac{1}{3}\left(g^{\mu \nu}-\frac{P^\mu P^\nu}{P^2}\right) \Theta^s(t) \\
=&\frac{1}{4m_N}\left[\Delta^\mu \Delta^\nu+\frac{\Delta^2}{3t}\left(P^\mu P^\nu-tg^{\mu \nu}\right)\right] A^s(t)\notag\\
&+\left[i \Delta^{\{\mu} \sigma^{\nu\} \rho} P_\rho+\frac{2i\sigma^{\rho\kappa}\Delta_\rho P_\kappa}{3t}\left(P^\mu P^\nu-tg^{\mu \nu}\right)\right]J^s(t)\ ,
\end{align}
respectively. The trace FF $\Theta^s(t)$ is defined as
\begin{align}\label{eq.nucleon_Theta}
    \left\langle N(p^\prime) \bar N(p)\left|\hat T^{\mu}_{\ \mu}(0)\right|0\right\rangle \equiv \bar u(p^\prime)v(p)\Theta^s(t)\ .
\end{align}
It is related to the GFFs $A^s(t), J^s(t), D^s(t)$ as: 
\begin{align}
    \Theta^s(t)=\frac{1}{4m_N}\left[-4p_N^2A^s(t)+2t J^s(t)-3tD^s(t)\right] ,
    \label{eq.nucleon_Theta_relation}
\end{align}
and it satisfies the following unitarity relation~\cite{Hoferichter:2023mgy}:
\begin{align}\label{eq.im_Theta}
    \operatorname{Im}\Theta^s(t)=-\frac{3p_\pi}{4p_N^2\sqrt{t}}\left(f^0_+(t)\right)^*\Theta^\pi(t)\ .
\end{align}
This can also be generalized by including $K\bar K$ intermediate states~\cite{Hoferichter:2023mgy}:
\begin{align}
    \operatorname{Im}\Theta^s(t)=-\frac{3}{4p_N^2\sqrt{t}}\left[p_\pi\left(f^0_+(t)\right)^*\Theta^\pi(t)\theta(t-t_\pi)+\frac{4}{3}p_K\left(h^0_+(t)\right)^* \Theta^K(t)\theta(t-t_K)\right] ,
\end{align}
where $h^0_+$ is the $S$-wave isospin-even $K\bar{K}\rightarrow N \bar{N}$ scattering amplitude.

Once the spectral functions of the nucleon GFFs are determined, the DRs can be immediately formulated for the GFFs.
In the timelike region, unitarity constraint implies that any FFs must vanish at infinite momentum transfer $t \rightarrow +\infty$~\cite{PhysRev.119.463}.
The exact behavior of the decaying power could be determined by the leading order pQCD~\cite{Tong:2021ctu, Tong:2022zax} and the Phragm\`en-Lindel{\"o}f theorem~\cite{Hohler:1983, Pacetti:2014jai}, which states that the spacelike asymptotic behavior of any FFs can be extended to any direction in the complex $t$ plane.
Therefore, the GFFs satisfy the unsubtracted DR (for simplicity, we omit the superscript ``$s$" or ``$N$" for the nucleon GFFs in the following): 
\begin{align}\label{eq.DR_AJTheta}
    (A,J,\Theta)(t) & =\frac{1}{\pi}\int^\infty_{t_\pi}\md t^\prime\frac{\operatorname{Im}(A,J,\Theta)(t^\prime)}{t^\prime-t}\ .
\end{align}
Thus, from the normalizations of the GFFs, one has the following sum rules:
\begin{align}\label{eq.sumrules_norm}
    (A,J,\Theta)(0) & =\frac{1}{\pi} \int_{t_\pi}^{\infty} \mathrm{d} t^{\prime} \frac{\operatorname{Im} (A,J,\Theta)\left(t^{\prime}\right)}{t^{\prime}}=\left(1,\frac{1}{2},m_N\right) .
\end{align}
However, these sum rules converge slowly, especially for $\Theta(t)$; see the weighted spectral functions for $(A,J,\Theta)(t)/t$ in Fig.~\ref{fig.ImGFF_nucleon}. A detailed discussion of sum rule saturation for nucleon GFFs appears in Appendix~\ref{app.sum rules}.
This phenomenon was discovered long ago in the dispersive analysis of the nucleon electromagnetic FFs~\cite{Belushkin:2006qa, Hoferichter:2016duk, Alarcon:2018irp, Lin:2021umz}.
Here, following the methodology in Refs.~\cite{Belushkin:2006qa, Hoferichter:2016duk, Alarcon:2018irp, Lin:2021umz}, we include additional effective zero-width poles with masses $m_{S,D}$ in the spectral functions, to simulate the contributions from possible high-lying excited meson states~\cite{Cao:2024zlf},
\begin{align}\label{eq.res.SD}
    \operatorname{Im} F(t)=\pi c_{S,D} m_{S,D}^2 \delta\left(t-m_{S,D}^2\right) ,
\end{align}
where $F\in\{A,J,\Theta\}$, and the subscripts ``$S$", ``$D$" denote $S$- and $D$-waves, respectively. 
The couplings $c_{S,D}$ are used to enforce the sum rules to be fulfilled. 
This formulation also allows for uncertainty estimates for the spectral densities by varying the effective pole masses $m_{S,D}$. 
We take the $m_D= 1.8^{+0.4}_{-0.3}~$GeV, which covers the mass range of $f_2(1565),f_2(1950)$ and $f_2(2010)$, and $m_S= 1.6^{+0.2}_{-0.1}~$GeV, covering the scalar resonances $f_0(1500)$ and $f_0(1710)$.
\begin{figure}[t]
    \centering
    \includegraphics[width=1\textwidth,angle=-0]{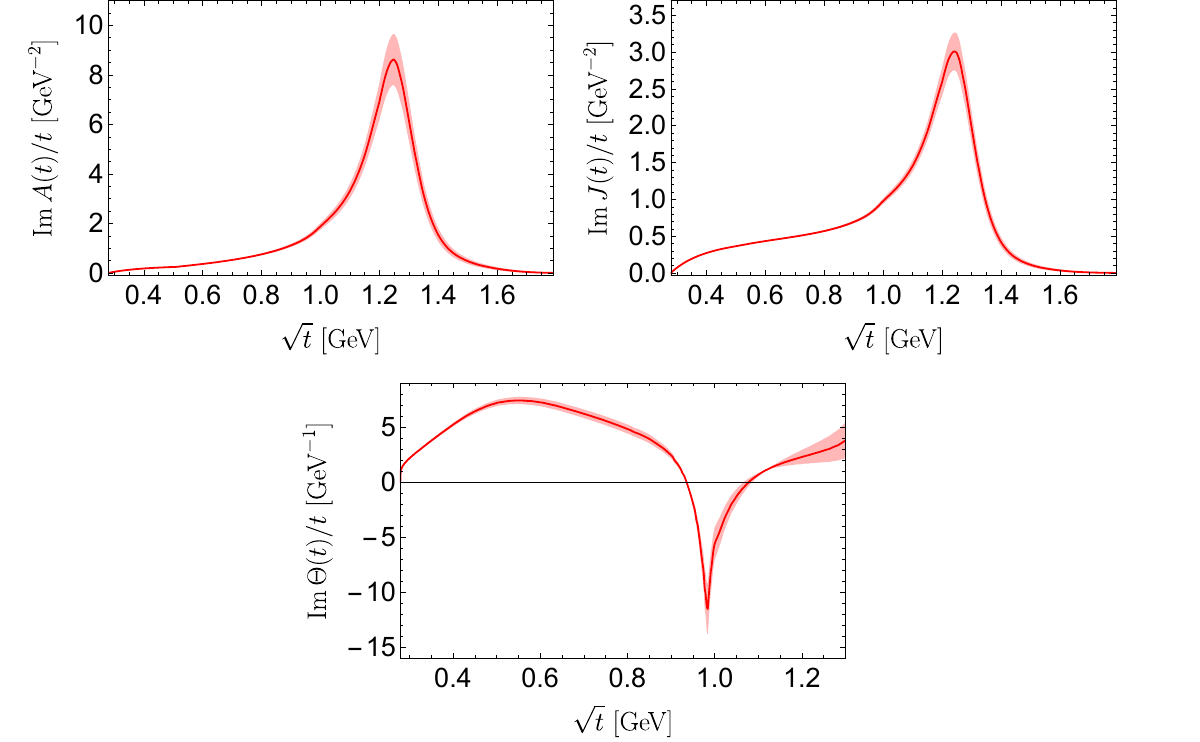} 
    \caption{Results of the nucleon weighted spectral functions for $A(t)/t,J(t)/t$ and $\Theta(t)/t$, respectively.} \label{fig.ImGFF_nucleon}
\end{figure}

\begin{figure}[t]
    \centering
    \includegraphics[width=1\textwidth,angle=-0]{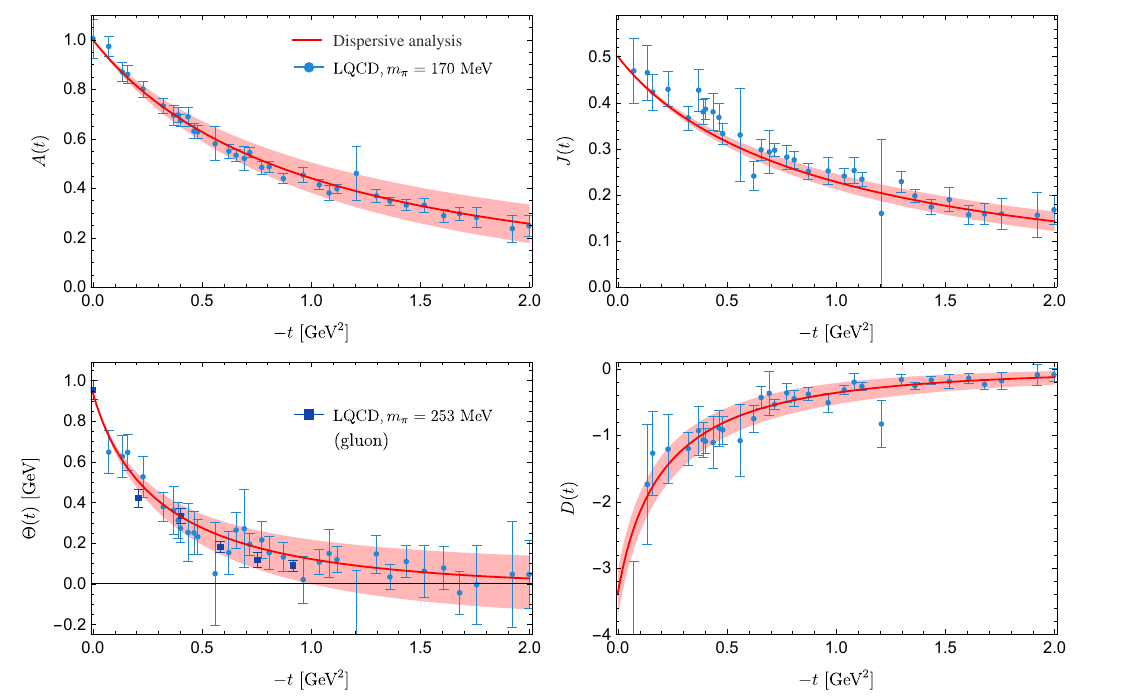} 
    \caption{The four total GFFs of the proton, computed based on the dispersive technique~\cite{Cao:2024zlf}. We also show the LQCD data at $m_\pi = 170~$MeV and $m_\pi = 253~$MeV taken from  Ref.~\cite{Hackett:2023rif} and Ref.~\cite{Wang:2024lrm}, respectively. The lattice results of $\Theta(t)$ at 170~MeV are obtained from a linear combination of the other three GFFs in Ref.~\cite{Hackett:2023rif}, with errors added in quadrature.} \label{fig.GFFN}
\end{figure}

Results of the DR analysis~\cite{Cao:2024zlf} are shown in Fig.~\ref{fig.GFFN}.
The input $\pi\pi/K\bar K \to N\bar N$ $S$-waves are taken from Roy-Steiner equation~\cite{Hite:1973pm} analysis in Ref.~\cite{Cao:2022zhn} (for other analyses, see Refs.~\cite{Hoferichter:2015hva, Hoferichter:2023mgy}. 
This method imposes general constraints on $\pi N$ scattering amplitudes including analyticity, unitarity, and crossing symmetry. The partial waves for $\pi\pi \rightarrow N \bar{N}$~\cite{Buettiker:2003pp, Descotes-Genon:2006sdr} are incorporated into a fully crossing-symmetric dispersive analysis, ensuring that the spectral function complies with all analytic $S$-matrix theory requirements and low-energy data constraints~\cite{Cao:2024zlf}. 
This approach has been used to derive model-independent results on nucleon properties, including scalar FFs~\cite{Hoferichter:2012wf}, the $\pi N$ $\sigma$-term~\cite{Hoferichter:2015dsa, RuizdeElvira:2017stg, Hoferichter:2023ptl}, electromagnetic FFs~\cite{Hoferichter:2016duk, Lin:2021umz}, and antisymmetric tensor FFs~\cite{Hoferichter:2018zwu}.  
For the $D$-wave, we adopt an input that differs slightly from that in Ref.~\cite{Hoferichter:2015hva}. 
The main difference lies in our use of the phase shift and inelasticity from Ref.~\cite{Bydzovsky:2016vdx}, which are consistent with those used in Ref.~\cite{Hoferichter:2015hva} up to 1.4~GeV and cover a larger energy range up to approximately 2~GeV. 
We have verified that the impact of the difference is almost negligible.

The main sources of uncertainties in our DR analysis arise from three aspects: the uncertainty from the meson GFFs, the $\pi\pi/K\bar K \to N\bar N$ partial waves, and the effective resonance masses. 
The first source of uncertainty has been addressed in Sec.~\ref{sec.3}, the second is provided by Ref.~\cite{Hoferichter:2015hva}, while the third is estimated by the relevant excited meson states as discussed in Eq.~\eqref{eq.res.SD}. A detailed error analysis can be found in Ref.~\cite{Cao:2024zlf}.

As can be seen from Fig.~\ref{fig.GFFN}, our results are quite close to the LQCD results of the gluonic trace anomaly GFF in Ref.~\cite{Wang:2024lrm}, which were obtained with an unphysical pion mass $m_\pi=253$~MeV. 
This demonstrates that the primary contribution to the nucleon trace GFF comes from gluons. 
Furthermore, another recent LQCD calculation in Ref.~\cite{Hackett:2023rif} has reported both the quark and gluon contributions to the GFFs $A, J$, and $D$, which can be combined to form the nucleon trace GFF via Eq.~\eqref{eq.nucleon_Theta_relation}. 
As reported in Refs.~\cite{Hackett:2023rif, Binosi:2025kpz}, the quark contribution to the three GFFs is consistently larger than the gluon contribution, though this difference remains modest. 
This seems to contradict the above statement about the gluon dominance of the trace GFF. 
However, there is no contradiction. The superficial discrepancy arises from that the calculations in Ref.~\cite{Hackett:2023rif,Binosi:2025kpz}\footnote{For the trace GFF as reported in Ref.~\cite{Wang:2024lrm}, this issue does not arise because they directly compute the scalar matrix element of the trace anomaly. Note that the scalar FF of the trace anomaly and the $\sigma$-term are scale-independent, respectively. This contrasts with the calculation of the scalar trace FF from the GFFs $A, J, D$ and $\bar{c}$.} did not consider the $\bar{c}_a$ GFF~\cite{Polyakov:2018zvc} ($a=q, g$ denotes the quark or gluon contribution).
They appear when discussing the quark and gluon GFFs separately, but sum up to zero, i.e., $\sum_{a}\bar{c}_a=0$, which ensures the conservation of the total energy-momentum, and thus do not contribute to the total GFFs discussed above.
When performing the quark-gluon decomposition of the trace GFF $\Theta(t)$, the $\bar{c}_a(t)$ term plays a significant role by accounting for the non-conservation of the separate quark and gluon parts of the EMT, leading to the conclusion that the trace GFF is predominantly driven by the gluon contribution when the $\bar c_a$ FFs are considered.
A detailed analysis can be found in Ref.~\cite{Lorce:2021xku}.

From the $D$-term FF in Fig.~\ref{fig.GFFN}, one gets the $D$-term as~\cite{Cao:2024zlf} 
\begin{align}
    D=-3.38^{+0.34}_{-0.35}\ ,
\end{align}
which satisfies the positivity bound~\cite{Gegelia:2021wnj}, $D \leq -0.20(2)$, and is in good agreement with the recent LQCD results, $-3.87(97)$ from the dipole fit and $-3.35(58)$ from the $z$-expansion fit in Ref.~\cite{Hackett:2023rif}.

\subsection{Spatial density profiles and mean squre radii}\label{sec.5}

Through the definition involving the EMT and by analogy to macroscopic continuum systems, various 3D static spatial densities within the proton can be defined in the Breit frame (BF)~\cite{Ernst:1960zza, Sachs:1962zzc}. 
However, the interpretation of probabilistic distributions of FFs in the BF is known to suffer from relativistic recoil corrections~\cite{Burkardt:2000za, Miller:2018ybm, Jaffe:2020ebz}.
Alternatively, uncontroversial 2D densities can be defined on the light front (LF)~\cite{Burkardt:2002hr, Miller:2010nz}, but they exhibit distortions that are difficult to reconcile with the picture of a classical system at rest~\cite{Miller:2007uy, Lorce:2020onh}. 
For in-depth discussions and reviews regarding the shape of light hadrons as relativistic QCD bound states, we refer to Refs.~\cite{Alexandrou:2012da, Gao:2021sml}. 
Various attempts to clarify the concept of relativistic spatial distributions have been recently undertaken~\cite{Epelbaum:2022fjc, Freese:2022fat, Li:2022hyf}.

One way to reconcile the BF and LF distributions is to adopt the Wigner phase-space distribution approach. 
This approach allows one to embrace the frame dependence of relativistic spatial distributions by relaxing the probabilistic density interpretation to a quasi-probabilistic one~\cite{Lorce:2017wkb, Lorce:2020onh}. 
In the probabilistic picture, the state is perfectly localized in position space and the expectation value of an operator $O$ is written as $\langle O\rangle=\int \mathrm{d}^3 R|\Psi(\boldsymbol{R})|^2 O(\boldsymbol{R})$ with $\Psi(\boldsymbol{R})$ being the position space wave packet. 
In the quasi-probabilistic picture, the expectation value is expressed as 
\begin{align}
    \langle O\rangle=\int \frac{\md^3 Q}{(2 \pi)^3} \mathrm{d}^3 R~\rho_{\Psi}(\boldsymbol{R}, \boldsymbol{Q}) O(\boldsymbol{R}, \boldsymbol{Q})\ ,
\end{align}
with $\rho_{\Psi}(\boldsymbol{R}, \boldsymbol{Q})$ being the Wigner distribution, a real-valued function that describes the localization of the system in phase space. 
Due to Heisenberg's uncertainty principle, Wigner distributions can be negative in some regions and therefore cannot be interpreted strictly as probability densities. 
The matrix element $O({\boldsymbol{R}, \boldsymbol{Q}})$ does not depend on the wave packet and can be interpreted as the expectation value in a state localized in the Wigner distribution sense around an average position $\boldsymbol{R}$ with an average momentum $\boldsymbol{Q}$.
Therefore, the key advantage of the phase-space approach is that it is unnecessary to explicitly isolate and subtract the wave-packet structure contribution by hand, which is irrelevant to structure analysis~\cite{Jaffe:2020ebz}. 

Another approach seeks to define three-dimensional (3D) densities using the instant form coordinates by taking the expectation value of a local current for a physical state and localizing the wave packet~\cite{Epelbaum:2022fjc}. 
By employing spherically symmetric wave packets in the zero-average-momentum frame (ZAMF) of the system, it has been demonstrated that 3D densities can be unambiguously defined for sharply localized wave packets. 
For discussions on various local spatial densities in the ZAMF, see Refs.~\cite{Panteleeva:2022khw, Panteleeva:2022uii, Panteleeva:2024vdw}.

Throughout this section, we mainly follow the conventions of Refs.~\cite{Polyakov:2018zvc, Lorce:2018egm}. 
The following discussion of 3D densities consistently adopts the phase-space perspective and also addresses spatial densities in the ZAMF. 
In the 3D BF (interpreted from the phase-space perspective as the average rest frame), the momenta of the incoming and outgoing hadrons are given by $p_\mu=(P_\mu-\Delta_\mu)/2$ and $p^{\prime}_\mu=(P_\mu+\Delta_\mu)/2$, such that $P^\mu=(2 E, \mathbf{0})$ (i.e., $\boldsymbol{P}=\mathbf{0}$) and $\Delta^\mu=(0, \boldsymbol{\Delta})$ while $t=-\boldsymbol{\Delta}^2$. 
Here, we consider only the spin-1/2 case, which is relevant for the proton. 
The nucleon GFFs $A(t), J(t), D(t)$ and $\Theta(t)$ can be related to the mass, scalar trace (also called dilatation), AM (also called spin) and radial (also called normal or longitudinal) force densities as~\cite{Polyakov:2002yz, Polyakov:2018zvc, Lorce:2017wkb, Lorce:2018egm, Kharzeev:2021qkd}:\footnote{Here the mass density is defined from the $00$ (i.e., energy) component of the EMT. We will use mass density and energy density interchangeably in the following.}
\begin{align}
    \rho_\text{Mass}(r)= &\ m_N \int \frac{\md^3 \boldsymbol{\Delta}}{(2 \pi)^3} e^{-i \boldsymbol{r}\cdot \boldsymbol{\Delta}}\left[A\left(t\right)-\frac{t}{4 m_N^2}\left[A\left(t\right)-2 J\left(t\right)+D\left(t\right)\right]\right] \label{eq.Edensity_BF}\nonumber\\
    = & \int \frac{\md^3 \boldsymbol{\Delta}}{(2 \pi)^3} e^{-i \boldsymbol{r}\cdot \boldsymbol{\Delta}}\left[\Theta\left(t\right)+\frac{t}{2 m_N}D(t)\right] ,\\
    \rho_\Theta(r)= &\ m_N \int \frac{\md^3 \boldsymbol{\Delta}}{(2 \pi)^3} e^{-i \boldsymbol{r}\cdot \boldsymbol{\Delta}}\left[A\left(t\right)-\frac{t}{4 m_N^2}\left[A\left(t\right)-2 J\left(t\right)+3D\left(t\right)\right]\right] \nonumber\\
    = & \int \frac{\md^3 \boldsymbol{\Delta}}{(2 \pi)^3} e^{-i \boldsymbol{r}\cdot \boldsymbol{\Delta}}\Theta\left(t\right) ,\\
    \rho_J(r)= &\ \int \frac{\md^3 \boldsymbol{\Delta}}{(2 \pi)^3} e^{-i \boldsymbol{r}\cdot \boldsymbol{\Delta}}\left[J\left(t\right)+\frac{2}{3} t \frac{\md}{\md t}J\left(t\right)\right] ,\\
    p_r(r)= &\ m_N\int \frac{\md^3 \boldsymbol{\Delta}}{(2 \pi)^3} e^{-i \boldsymbol{r}\cdot \boldsymbol{\Delta}}\left[-\frac{1}{r^2}\frac{1}{\sqrt{t}m_N^2}\frac{\md}{\md t}t^{\frac{3}{2}}D(t)\right]\equiv p(r)+\frac{2}{3}s(r)\ ,\\
    p(r)= &\ \frac{1}{6 m_N} \frac{1}{r^2} \frac{\md}{\md r} r^2 \frac{\md}{\md r} \tilde{D}(r)\ ,\quad s(r)=-\frac{1}{4 m_N} r \frac{\md}{\md r} \frac{1}{r} \frac{\md}{\md r} \tilde{D}(r)\ ,
\end{align}
where $r=|\boldsymbol{r}|$, $\tilde{D}(r)=\int \frac{\md^3 \boldsymbol{\Delta}}{(2 \pi)^3} e^{-i \boldsymbol{r}\cdot \boldsymbol{\Delta}} D\left(t\right)$, and $p(r)$ and $s(r)$ correspond to the pressure and shear force densities, respectively. 
We can also define the tangential force density as $p_t(r)\equiv p(r)-s(r)/3$. 
Following the standard convention, we have neglected the contribution of the quadrupole term in the AM density. 
The monopole and quadrupole AM distributions are not independent but are interrelated as specified in Refs.~\cite{Lorce:2017wkb, Schweitzer:2019kkd}, i.e., $\rho_{J, \text {quad}}(r)=-3\rho_J(r)/2$. 
The distribution $\rho_J(r)$ given above corresponds to the monopole contribution.

In the ZAMF, we consider the spatial energy density $\rho_\text{Mass}^\text{ZAMF}(r)$ as an example. The connection between the GFFs and finite mass density $\rho_\text{Mass}^\text{ZAMF}(r)$ is given by~\cite{Panteleeva:2022uii}\footnote{In our convention, $\rho_\text{Mass}^\text{ZAMF}(r)$ corresponds to the expression $t_{00}(r)/(4\pi N_{\phi,R})$ in Ref.~\cite{Panteleeva:2022uii}.}
\begin{align}
    \rho_\text{Mass}^\text{ZAMF}(r)=\frac{m_N}{4\pi r}\int^\infty_0 \md \Delta~\Delta \sin\left(\Delta r\right)\int^1_{-1}\md \alpha~A\!\left((\alpha^2-1)\Delta^2\right) ,
\end{align}
with $\Delta=|\boldsymbol{\Delta}|$.
For comparison, we also provide the definition of the ``naive'' energy density, which corresponds to taking the limit $m_N \to \infty$ (infinite mass limit) in the BF expression~Eq.~\eqref{eq.Edensity_BF},
\begin{align}
    \rho_\text{Mass}^\text{naive}(r)= & m_N \int \frac{\md^3 \boldsymbol{\Delta}}{(2 \pi)^3} e^{-i \boldsymbol{r}\cdot \boldsymbol{\Delta}}A\left(t\right) .
\end{align}

\begin{figure}[t]
    \centering
    \includegraphics[width=1\textwidth,angle=-0]{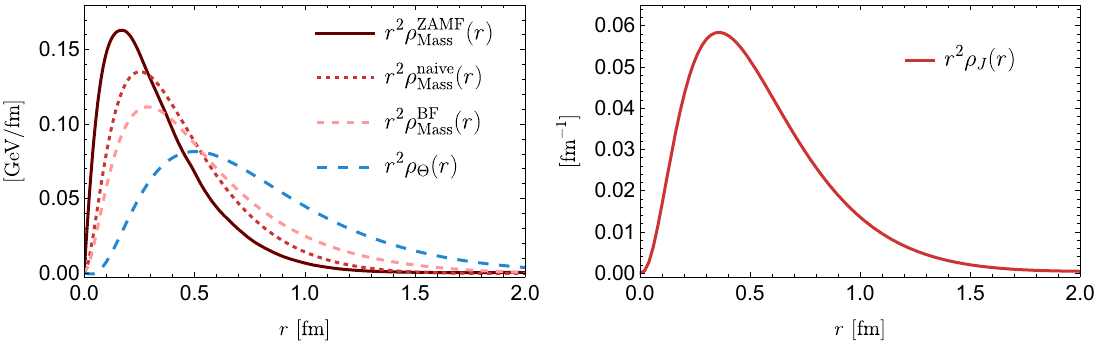}
    \caption{Left panel: the nucleon mass distributions $r^2 \rho_\text{Mass}^\text{ZAMF}(r)$ in the ZAMF, $r^2 \rho_\text{Mass}^\text{naive}(r)$ in the static limit, $r^2 \rho_\text{Mass}^\text{BF}(r)$ in the BF, with $\rho_\text{Mass}^\text{BF}(r)$ the one in Eq.~\eqref{eq.Edensity_BF}, and the trace density distribution $r^2 \rho_\Theta(r)$ in the BF, respectively.
    Right panel: the nucleon AM distribution in the BF.} \label{fig.EMJ_density}
\end{figure}

\begin{figure}[t]
    \centering
    \includegraphics[width=1\textwidth,angle=-0]{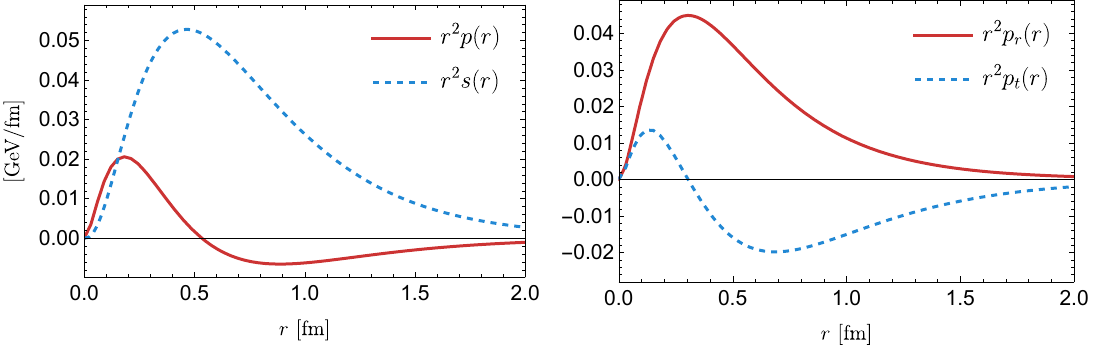}
    \caption{Left panel: the pressure and shear force distributions inside a nucleon. 
    Right panel: the radial and tangential force distributions inside a nucleon.} \label{fig.pressure3D}
\end{figure}

A stable numerical Fourier transformation requires knowledge of the complete $Q^2$-dependence of FFs, where $Q^2=-t$. 
The large $Q^2$ behavior of the spacelike nucleon GFFs is determined by the QCD hard scattering mechanism, up to logarithmic corrections, in Refs.~\cite{Tanaka:2018wea, Tong:2021ctu, Tong:2022zax, Liu:2024vkj} to be $A(Q^2) \sim J(Q^2) \sim \Theta(Q^2) \sim 1/Q^4$ and $D(Q^2) \sim 1/Q^6$.
With the above asymptotic behavior of the GFFs, Eq.~\eqref{eq.DR_AJTheta} implies the superconvergence relations; see Appendix~\ref{app.sum rules}.
Unfortunately, numerical values of the left-hand side of the above relations highly depend on the high-energy tails of the spectral functions, which are poorly known and present significant challenges for an accurate description. 
Therefore, we adopt a $z$-expansion~\cite{Hill:2010yb} method to extrapolate the GFFs in the spacelike region (for more details, see Appendix~\ref{app.z-exp}). 
So far, all currently known calculations of spatial densities corresponding to GFFs, in particular at short distances, are more or less model-dependent, as we lack model-independent GFFs spanning in the whole range of $t$ from $t=0$ to $\infty$.
Here, we use the central values of the GFFs derived in Ref.~\cite{Cao:2024zlf}, as reviewed in Sec.~\ref{sec.4}, as inputs.
These distributions are shown in Figs.~\ref{fig.EMJ_density} and~\ref{fig.pressure3D}.

The 3D BF densities yield the mass, scalar trace, angular momentum and mechanical root-mean-square (rms) radii, $r_i\equiv \sqrt{\left\langle r_i^2\right\rangle}$ ($i=$Mass, $\Theta,J$, Mech), of the proton~\cite{Polyakov:2018zvc} as~\cite{Cao:2024zlf}:\footnote{Here the ``mass radius'' $r_\text{Mass}$ is defined as the radius derived from the energy density; see, e.g., Ref.~\cite{Ji:2021mtz}. The radius of the trace density distribution is also called a ``scalar radius'' in the literature.}
\begin{align}
    \left\langle r_\text{Mass}^2\right\rangle= &\ \frac{\int \md^3 \boldsymbol{r}~r^2 \rho_\text{Mass}(r)}{\int \md^3 \boldsymbol{r}~\rho_\text{Mass}(r)}=6 A^\prime(0)-\frac{3 D}{2 m_N^2}=\left(0.70^{+0.03}_{-0.04}~\text{fm}\right)^2\ ,\\
    \left\langle r_\Theta^2\right\rangle= &\ \frac{\int \md^3 \boldsymbol{r}~r^2 \rho_\Theta(r)}{\int \md^3 \boldsymbol{r}~\rho_\Theta(r)}=6 A^\prime(0)-\frac{9 D}{2 m_N^2}=\left(0.97^{+0.03}_{-0.03}~\text{fm}\right)^2\ ,\\
    \left\langle r_J^2\right\rangle= &\ \frac{\int \md^3 \boldsymbol{r}~r^2 \rho_J(r)}{\int \md^3 \boldsymbol{r}~\rho_J(r)}=20 J^\prime(0)=\left(0.70^{+0.02}_{-0.02}~\text{fm}\right)^2\ ,\\
    \left\langle r_\text{Mech}^2\right\rangle= &\ \frac{\int \md^3 \boldsymbol{r}~r^2 p_r(r)}{\int \md^3 \boldsymbol{r}~p_r(r)}=\frac{6 D}{\int_{-\infty}^0 \md t~D(t)}=\left(0.72^{+0.09}_{-0.08}~\text{fm}\right)^2\ .
\end{align}
The mass mean-square radius in the ZAMF and the expression for ``naive" mass mean-square radius can be written as
\begin{align}
    \left\langle r_\text{Mass}^2\right\rangle^\text{ZAMF}= &\ \frac{\int \md^3 \boldsymbol{r}~r^2 \rho_\text{Mass}^\text{ZAMF}(r)}{\int \md^3 \boldsymbol{r}~\rho_\text{Mass}^\text{ZAMF}(r)}=4 A^\prime(0)=\left(0.42^{+0.02}_{-0.02}~\text{fm}\right)^2\ ,\\
    \left\langle r_\text{Mass}^2\right\rangle^\text{naive}= &\ \frac{\int \md^3 \boldsymbol{r}~r^2 \rho_\text{Mass}^\text{naive}(r)}{\int \md^3 \boldsymbol{r}~\rho_\text{Mass}^\text{naive}(r)}=6 A^\prime(0)=\left(0.52^{+0.02}_{-0.03}~\text{fm}\right)^2\ .
\end{align}
A smaller mean-square radius, $4 A^\prime(0)$, compared to the ``naive'' result, $6 A^\prime(0)$, has been obtained previously using the LF infinite-momentum frame (IMF) formalism~\cite{Miller:2018ybm}.
Intuitively, the 3D energy density under discussion in the ZAMF can be understood as the 2D transverse energy density rotated through all possible directions to fill a 3D volume and then averaged~\cite{Carlson:2022eps}.
Clearly, the complete representation of a 3D object can be reconstructed by combining all possible 2D projections.

Recall that the BF allows one to define 3D distributions for $\boldsymbol{P}=\mathbf{0}$. 
If we consider the scenario where $\boldsymbol{P} \neq \mathbf{0}$, we find that the distributions we can define are inherently 2D. 
We define the elastic frame (EF) by the condition $\boldsymbol{P} \cdot \boldsymbol{\Delta}=0$. 
These constitute a class of frames characterized by zero energy transfer to the system, i.e., $\Delta^0=0$. 
The BF appears as a particular element of this class. 
The particular case $P_z=0$ is simple and corresponds to the BF, for which we can obtain the following combinations~\cite{Lorce:2017wkb, Lorce:2018egm}:\footnote{Following the convention in Ref.~\cite{Lorce:2018egm}, we will refer to the special case of EF where $P_z=0$ simply as EF if no confusion arises.}
\begin{align}
    \sigma_\text{Mass}(b)= &\ m_N \int \frac{\md^2 \boldsymbol{\Delta}_\perp}{(2 \pi)^2} e^{-i \boldsymbol{b}_\perp\cdot \boldsymbol{\Delta}_\perp}\left[A\left(t\right)-\frac{t}{4 m_N^2}\left[A\left(t\right)-2 J\left(t\right)+D\left(t\right)\right]\right] \nonumber\\
    = & \int \frac{\md^2 \boldsymbol{\Delta}_\perp}{(2 \pi)^2} e^{-i \boldsymbol{b}_\perp\cdot \boldsymbol{\Delta}_\perp}\left[\Theta\left(t\right)+\frac{t}{2 m_N}D(t)\right] ,\label{eq.Edensity_EF}\\
    \sigma_\Theta(b)= &\ m_N \int \frac{\md^2 \boldsymbol{\Delta}_\perp}{(2 \pi)^2} e^{-i \boldsymbol{b}_\perp\cdot \boldsymbol{\Delta}_\perp}\left[A\left(t\right)-\frac{t}{4 m_N^2}\left[A\left(t\right)-2 J\left(t\right)+3D\left(t\right)\right]\right] \nonumber\\
    = & \int \frac{\md^2 \boldsymbol{\Delta}_\perp}{(2 \pi)^2} e^{-i \boldsymbol{b}_\perp\cdot \boldsymbol{\Delta}_\perp}\Theta\left(t\right) ,\\
    \sigma_J(b)= &\ \int \frac{\md^2 \boldsymbol{\Delta}_\perp}{(2 \pi)^2} e^{-i \boldsymbol{b}_\perp\cdot \boldsymbol{\Delta}_\perp}\left[J\left(t\right)+t \frac{\md}{\md t}J\left(t\right)\right]=\sigma_{J, \text{mono}}(b)+\sigma_{J, \text{quad}}(b)\ ,\\
    \sigma_{J, \text{mono}}(b)= &\ \frac{1}{2}\int \frac{\md^2 \boldsymbol{\Delta}_\perp}{(2 \pi)^2} e^{-i \boldsymbol{b}_\perp\cdot \boldsymbol{\Delta}_\perp}\left[J\left(t\right)+\frac{2}{3}t \frac{\md}{\md t}J\left(t\right)\right] ,\\ 
    \sigma_{J, \text{quad}}(b)= &\ \int \frac{\md^2 \boldsymbol{\Delta}_\perp}{(2 \pi)^2} e^{-i \boldsymbol{b}_\perp\cdot \boldsymbol{\Delta}_\perp}\frac{1}{3}t \frac{\md}{\md t}J\left(t\right) ,\\
    \sigma_r(b)= &\ m_N\int \frac{\md^2 \boldsymbol{\Delta}_\perp}{(2 \pi)^2} e^{-i \boldsymbol{b}_\perp\cdot \boldsymbol{\Delta}_\perp}\left[-\frac{1}{2 b^2}\frac{1}{m_N^2}\frac{\md}{\md t}t^{\frac{3}{2}}D(t)\right]\equiv \sigma(b)+\frac{1}{2}\Pi(b)\ ,\\
    \sigma_t(b)\equiv &\ \sigma(b)-\frac{1}{2}\Pi(b)\ ,\\ 
    \sigma(b)= &\ \frac{1}{8 m_N} \frac{1}{b} \frac{\md}{\md b} b \frac{\md}{\md b} \overline{D}(b)\ ,\quad \Pi(b)=-\frac{1}{4 m_N} b \frac{\md}{\md b} \frac{1}{b} \frac{\md}{\md b} \overline{D}(b)\ , \notag
\end{align} 
where $b=|\boldsymbol{b}_\perp|$ and $\overline{D}(b)=\int \frac{\md^2 \boldsymbol{\Delta}_\perp}{(2 \pi)^2} e^{-i \boldsymbol{b}_\perp\cdot \boldsymbol{\Delta}_\perp} D\left(t\right)$. 
Note that the asymptotic large-$b$ behaviour can be determined by the infrared physics contained in unitarity relations~\cite{Broniowski:2025ctl}.

\begin{figure}[t]
    \centering
    \includegraphics[width=1\textwidth,angle=-0]{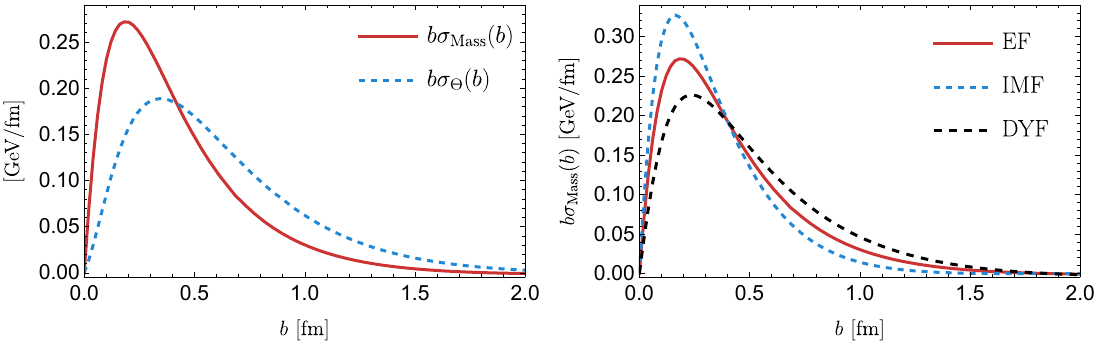}
    \caption{Left panel: the mass and scalar trace densities of a nucleon in the 2D EF, respectively. 
    Right panel: the mass densities of a nucleon in the 2D EF (with $P_z=0$), IMF and LF DYF, respectively.} \label{fig.EM_2Ddensity}
\end{figure}

\begin{figure}[t]
    \centering
    \includegraphics[width=1\textwidth,angle=-0]{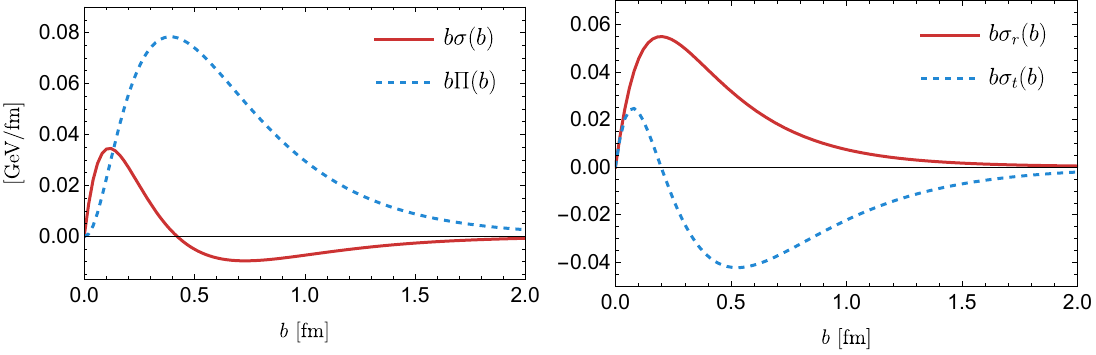}
    \caption{Left panel: the pressure and shear forces of a nucleon in the 2D EF. 
    Right panel: the radial and tangential forces of a nucleon in the 2D EF.} \label{fig.pressure2D}
\end{figure}

\begin{figure}[t]
    \centering
    \includegraphics[width=.5\textwidth,angle=-0]{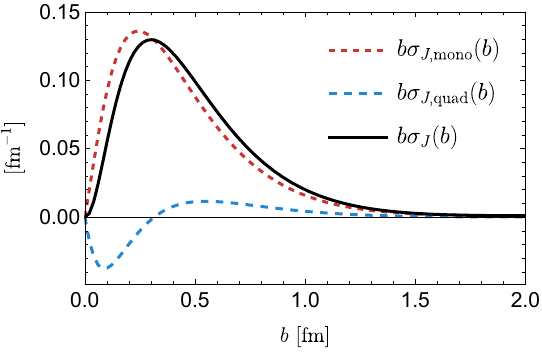}
    \caption{The monopole, quadrupole and total AM densities of a nucleon in the 2D EF.} \label{fig.J_2Ddensity}
\end{figure}

In Figs.~\ref{fig.EM_2Ddensity},~\ref{fig.pressure2D} and~\ref{fig.J_2Ddensity}, we plot the above-mentioned 2D instant EF distributions as functions of the impact parameter $b$.

The IMF is a special case of the EF obtained by taking the limit $P_z\to \infty$. 
We can write the mass density~\cite{Lorce:2018egm},
\begin{align}
    \sigma_\text{Mass}^\text{IMF}(b)=m_N \int \frac{\md^2 \boldsymbol{\Delta}_\perp}{(2 \pi)^2} e^{-i \boldsymbol{b}_\perp\cdot \boldsymbol{\Delta}_\perp}A\left(t\right) .
\end{align}
Moreover, all 2D distributions except the mass density are identical in both the instant EF ($P_z=0$) and IMF ($P_z\to \infty$)~\cite{Lorce:2017wkb, Lorce:2018egm}. 
This can be understood from the fact that kinetic energy increases with $P_z$ while the intrinsic contributions from trace anomaly, AM and pressure forces remain constant. 
We illustrate the mass density in the IMF in Fig.~\ref{fig.EM_2Ddensity}.

In the LF formalism, which amounts to adopting the point of view of a massless observer, the subgroup of Lorentz transformations associated with the transverse plane is Galilean~\cite{Susskind:1967rg}. The LF Drell-Yan frame (DYF) is defined by $\Delta^+=({\Delta^0+\Delta^3})/2=0$. 
These LF distributions in the DYF coincide with the corresponding instant form distributions in the EF~\cite{Lorce:2017wkb, Lorce:2018egm}.
Moreover, the instant form and the LF form coincide in the IMF where $P_z \rightarrow \infty$.
Since the 2D distributions we consider do not depend on $P_z$, they should be the same in both the instant form and the LF form. 
It is also not surprising that the energy densities defined in the instant BF form~\eqref{eq.Edensity_EF} and LF DYF form~\cite{Lorce:2018egm}
\begin{align}
    \sigma_\text{Mass}^\text{DYF}(b)= &\ m_N \int \frac{\md^2 \boldsymbol{\Delta}_\perp}{(2 \pi)^2} e^{-i \boldsymbol{b}_\perp\cdot \boldsymbol{\Delta}_\perp}\left[A\left(t\right)-\frac{t}{2 m_N^2}\left[A\left(t\right)-2 J\left(t\right)+D\left(t\right)\right]\right] \nonumber\\
    = &\ \int \frac{\md^2 \boldsymbol{\Delta}_\perp}{(2 \pi)^2} e^{-i \boldsymbol{b}_\perp\cdot \boldsymbol{\Delta}_\perp}\left[2\Theta\left(t\right)+\frac{t}{m_N}D(t)-m_N A(t)\right] ,
\end{align}
differ because they are simply related to different components of the EMT. 
The latter is also illustrated in Fig.~\ref{fig.EM_2Ddensity}.
It is noted that the total LF energy is given by $\int \mathrm{d}^2 b_{\perp} \sigma_\text{Mass}^\text{DYF}(b)=m_N$ instead of $m_N/2$ from the standard definition of the LF components of EMT~\cite{Lorce:2018egm}, i.e., $\sigma_\text{Mass}^\text{DYF}(b)=2 \mu(b)$, where $\mu(b)$ is the energy density defined in Ref.~\cite{Lorce:2018egm}. 

\subsection{Matching dispersion relation to chiral perturbation theory}\label{sec.6}

To systematically study scattering processes induced in the presence of gravitational external field in the low-energy domain, one can construct the effective chiral Lagrangian for pions and nucleons in curved spacetime. 
The corresponding mesonic $\mathcal{O}(p^4)$ Lagrangian for pNGBs has been constructed in Ref.~\cite{Gasser:1984gg}, and the pion GFFs have also been obtained based on the chiral Lagrangian.  
The Lagrangian contains three additional terms ($\propto L_{11},L_{12},L_{13}$) compared to the flat space-time case~\cite{Gasser:1984gg}.   
These results are then generalized to the isospin violation case in Ref.~\cite{Kubis:1999db}.
Although these additional terms vanish in flat spacetime, they contribute to the EMT of pions in Minkowski spacetime and hence to the pion GFFs. 

The nucleon sector was first investigated in heavy baryon ChPT in Refs.~\cite{Belitsky:2002jp, Diehl:2006ya}. 
In Ref.~\cite{Alharazin:2020yjv}, the chiral effective Lagrangian for nucleons has been extended to curved spacetime up to NLO, i.e., $\mathcal{O}(p^2)$, by introducing two additional terms ($\propto c_8,c_9$), and it was used to obtain the nucleon GFFs up to $\mathcal{O}(p^4)$. 
Recently, this approach has been generalized to quantities such as the $\rho$ meson GFFs~\cite{Epelbaum:2021ahi}, the $\Delta$ resonance GFFs~\cite{Alharazin:2022wjj}, the $p \to \Delta^+$ transition GFFs~\cite{Alharazin:2023zzc}, the deuteron GFFs~\cite{Panteleeva:2024abz}, and the nucleon GFFs up to $\mathcal{O}(p^4)$ with the $\Delta$ resonances as explicit degrees of freedom in the small-scale expansion scheme~\cite{Alharazin:2023uhr}.
However, the new LECs, $c_8$ and $c_9$, have not yet been determined. 
Only an estimate for $c_8$ exists based on positivity bound~\cite{Gegelia:2021wnj}, while $c_9$ remains completely unknown.
With the growing prevalence of gravity-induced ChPT calculations, there is an urgent need for precise, model-independent determinations of the LECs $c_8$ and $c_9$. 
In this subsection, we address this issue by matching the nucleon GFFs presented in Sec.~\ref{sec.4} to baryon ChPT (BChPT).

\subsubsection{Matching at the next-to-leading order}

We follow the notation in Ref.~\cite{Alharazin:2020yjv}. The $\mathcal{O}(p^2)$ chiral action contains one term proportional to the Ricci tensor $R^{\mu\nu}$ and one term proportional to the Ricci scalar $R = g_{\mu\nu} R^{\mu\nu}$~\cite{Alharazin:2020yjv}:
\begin{align}
    \mathcal{S}_{\pi \mathrm{N}}^{(2)}=\int \md^4 x \sqrt{|g|}\left\{\frac{c_8}{8} R \bar{\psi} \psi+\frac{i c_9}{m_N} R^{\mu \nu}\left(\bar{\psi} e_\mu^a \gamma_a \nabla_\nu \psi-\nabla_\nu \bar{\psi} e_\mu^a \gamma_a \psi\right)\right\} .
\end{align}
Both $R$ and $R^{\mu\nu}$ are counted as $\mathcal{O}(p^2)$~\cite{Donoghue:1991qv}.
The LECs $c_8$ and $c_9$ can be constrained by the GFFs of the nucleon and are universal like all other LECs---the same constants appear in various low-energy processes probed by external gravitaional field.

The tree-level contributions up to $\mathcal{O}(p^2)$ to the GFFs are expressed as follows,
\begin{align}
    A^{(2)}(t) & =1-\frac{2 c_9}{m_N} t\ , \\
    J^{(2)}(t) & =\frac{1}{2}-\frac{c_9}{m_N} t\ , \\
    D^{(2)}(t) & =c_8 m_N\ .
\end{align}
Then we can easily obtain the following chiral $\mathcal{O}(p^2)$ expression for the $D$-term,
\begin{align}
    D=c_8 m_N\ ,
\end{align}
and for the slopes at $t=0$, 
\begin{align}
    \dot{A}(0)=2\dot{J}(0)=-\frac{2c_9}{m_N}\ .
\end{align}
The dispersive results for the slopes of GFFs $\dot{A}(0), \dot{J}(0)$ and the $D$-term can be easily propagated to the LECs $c_8$ and $c_9$, and we obtain
\begin{align}
    c_8=-3.60^{+0.37}_{-0.38}~\mathrm{GeV}^{-1}\ ,\quad c_9=-0.58^{+0.03}_{-0.03}~\mathrm{GeV}^{-1}\ .
\end{align}

\subsubsection{Matching at the next-to-next-to-leading order}

\begin{figure}[t]
    \centering
    \includegraphics[width=1\textwidth,angle=-0]{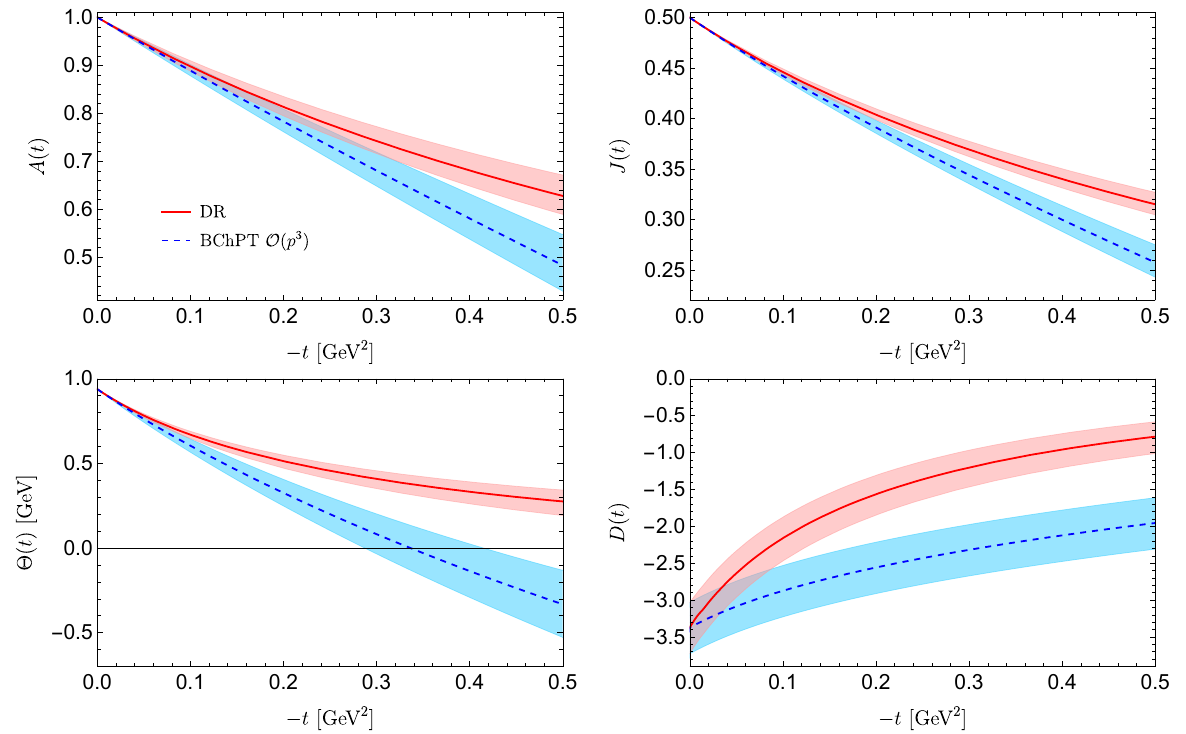}
    \caption{Comparision between the dispersive results of the nucleon GFFs and those obtained from $\mathcal{O}(p^3)$ BChPT.} \label{fig.GFFN_chpt}
\end{figure}
The $\mathcal{O}(p^3)$, i.e., next-to-next-to-leading order (NNLO), chiral action for nucleons containing the Riemann tensor $R^{\mu\nu\alpha\beta}$, the Ricci tensor and the Ricci scalar has five new terms, but only one of them contributes to nucleon GFFs~\cite{Alharazin:2023uhr},
\begin{align}
    \mathcal{S}_{\pi \mathrm{N}}^{(3)}=\int \md^4 x \sqrt{|g|}\left\{i \tilde{d}_{g 4} \nabla_\beta R^{\mu \nu \alpha \beta}\left(\bar{\psi}\sigma_{\mu \nu} e_\mu^a \gamma_a \nabla_\nu \psi-\nabla_\nu \bar{\psi}\sigma_{\mu \nu} e_\mu^a \gamma_a \psi\right)\right\} .
\end{align}
It gives an $\mathcal{O}(p^3)$ tree-level contribution to the GFF $J(t)$,
\begin{align}
    J^{(3)}(t) \supset 2 \tilde{d}_{g 4} m_N t\ .
\end{align}
By incorporating the leading one-loop contributions at NNLO~\cite{Alharazin:2023uhr}, the values of the slopes of the GFFs $\dot{A}(0), \dot{J}(0)$, and the $D$-term derived from DRs straightforwardly lead to values of the LECs $c_8$, $c_9$, and $\tilde{d}_{g 4}$,
\begin{align}
    c_8=-4.28^{+0.37}_{-0.38}~\mathrm{GeV}^{-1}\ ,\quad c_9=-0.68^{+0.06}_{-0.05}~\mathrm{GeV}^{-1}\ ,\quad \tilde{d}_{g 4}=-0.04^{+0.01}_{-0.02}~\mathrm{GeV}^{-3}\ ,
\end{align}
with the renormalization scale chosen as $\mu=m_N$.

In Fig.~\ref{fig.GFFN_chpt}, we compare the predictions from NNLO BChPT to the dispersive results of the nucleon GFFs up to $-t=0.5~\mathrm{GeV}^2$. 
The NNLO BChPT predictions for $A(t)$ and $J(t)$ are in better agreement with the dispersive results than those for $\Theta(t)$ and $D(t)$. 
This can be understood as follows: while both $A(t)$ and $J(t)$ depend only on the $D$-wave $\pi\pi$, $\Theta(t)$ and $D(t)$ also depend on the $S$-wave $\pi\pi$; the contribution from the $\sigma$ meson in the $S$-wave $\pi\pi$ scattering amplitude, which has a much smaller mass than the $f_2(1270)$ resonance, is not effectively captured in the NNLO BChPT prediction but is contained in the DR analysis.

\section{Summary and outlook}\label{sec.7}

In this work, as an extension of our analysis of the pion and nucleon GFFs in Ref.~\cite{Cao:2024zlf} using dispersive techniques, we have comprehensively investigated the total GFFs of the pion, kaon and nucleon. 
Our study provides new insights into the pion mass dependence of the GFFs, and the spatial distributions of mass, energy, angular momentum, and shear forces within nucleons. 
In the following, we summarize the specific results obtained in this work and also briefly mention those obtained in Ref.~\cite{Cao:2024zlf}.
\begin{itemize}
    \item[$1.$] The GFFs of pNGBs at small momentum transfers can be described by ChPT. 
    Utilizing precise low-energy scattering phase shift data for $\pi\pi$ and $K\bar{K}$ scatterings, we constructed the GFFs of pions and kaons using the Muskhelishvili-Omn\`es formalism, matched to ChPT at NLO at small momentum transfers~\cite{Cao:2024zlf}. 
    The model-independent results of the pion GFFs obtained in Ref.~\cite{Cao:2024zlf} agree well with the latest lattice QCD calculations near the physical pion mass~\cite{Hackett:2023nkr}.
    The kaon GFFs are predicted in the same way. 
    
    \item[$2.$] We extended the results for the pion to three unphysical pion masses, $m_{\pi}=239, 283$, and $391$~MeV, for which Roy-equation analysis results for the phase shift data are available~\cite{Cao:2023ntr, Rodas:2023nec}. 
    We found that the pion GFFs exhibit marginal dependence on the pion mass for $m_{\pi}$ up to 283~MeV, when the $\sigma$ resonance remains a resonance pole.  
    When $m_\pi=391$~MeV, the $\sigma$ has already become a bound state, and significant changes occur in the $\Theta^\pi$ and $D^\pi$ GFFs, which depend on the isospin-zero $S$-wave phase shift where the $\sigma$ meson resides. 
    
    \item[$3.$] The nucleon GFFs at the physical pion mass have also been obtained in a model-independent way using the DR approach in Ref.~\cite{Cao:2024zlf}, which also reported the $D$-term of the nucleon to be $D=-3.38_{-0.35}^{+0.34}$. 
    These results provide an effective means of probing the internal nucleon distributions, such as mass, scalar trace density, angular momentum, and stress. 
    Direct experimental measurements of these distributions are not feasible in a model-independent manner, thus the results from the dispersive approach are very useful. 
    We have explored three-dimensional spatial density distributions and two-dimensional transverse density distributions using the Wigner phase-space quasi-probability form, the zero-angular momentum frame form, and also the light-front form. 
    
    \item[$4.$] ChPT in the presence of an external gravitational field systematically describes various low-energy interactions induced by gravity. 
    By matching the $D$-term and the slopes of the GFFs at $t=0$ for the nucleon obtained using DR techniques to those from BChPT, we determined the values of new LECs in the BChPT Lagrangian in curved spacetime for the first time. 
    At NLO, we obtained $c_8=-3.60^{+0.37}_{-0.38}~\mathrm{GeV}^{-1}$ and $c_9=-0.58^{+0.03}_{-0.03}~\mathrm{GeV}^{-1}$, and at NNLO, we found $c_8=-4.28^{+0.37}_{-0.38}~\mathrm{GeV}^{-1}\ ,\quad c_9=-0.68^{+0.06}_{-0.05}~\mathrm{GeV}^{-1}$ and $\tilde{d}_{g 4}=-0.04^{+0.01}_{-0.02}~\mathrm{GeV}^{-3}$ at the renormalization scale $\mu=m_N$, where $c_{8,9}$ are LECs at in the NLO Lagrangian and $\tilde{d}_{g 4}$ is a new LEC at NNLO. 
    These LECs can be used to compute other low-energy gravity-induced scattering processes and are applicable to extracting physical information from future experiments and lattice studies. 
\end{itemize}

In the future, we plan to use a similar approach to investigate the GFFs of hyperons and their mass radii. 
Additionally, we plan to systematically investigate the pion mass dependence of nucleon GFFs, analogous to recent studies on the pion mass dependence of nucleon electromagnetic form factors~\cite{Alvarado:2023loi}.

\bmhead{Acknowledgements}
We would like to thank Jambul Gegelia for helpful discussions and the authors of Ref.~\cite{Hackett:2023rif} for providing their lattice data for Fig.~\ref{fig.GFFN}. We also thank Herzallah Alharazin for providing the complete expressions of the gravitational form factors of the nucleon from Ref.~\cite{Alharazin:2023uhr}.
XHC acknowledges the school of physics and electronics, Hunan University, for the very kind hospitality during his visit. DLY appreciates the support of Peng Huan-Wu visiting professorship and the hospitality of Institute of Theoretical Physics at Chinese Academy of Sciences (CAS). 
This work was supported in part by CAS under Grant No.~YSBR-101; by the National Natural Science Foundation of China under Grants No.~12347120, No.~12275076, No.~12335002, No.~12375078, No.~11905258, No.~12125507, No.~12361141819, and No.~12447101; by the Postdoctoral Fellowship Program of China Postdoctoral Science Foundation under Grants No.~GZC20232773 and No.~2023M74360; by the National Key R\&D Program of China under Grant No.~2023YFA1606703; and by the Science Fund for Distinguished Young Scholars of Hunan Province under Grant No.~2024JJ2007.

\bmhead{Data availability statement}
All the results presented in this work can be retrieved from the cited references, and we request the interested reader to consult those references for detail.

\begin{appendices}

\numberwithin{equation}{section}
\renewcommand{\theequation}{\Alph{section}\arabic{equation}}
\numberwithin{table}{section}
\renewcommand{\thetable}{\Alph{section}\arabic{table}}

\section{Sum rules of gravitational form factors}\label{app.sum rules}

Converting dispersion relations into sum rules relies on low- or high-energy expansions. Expanding both sides of the dispersion relations~\eqref{eq.DR_AJTheta} in a Taylor series around $t\to\infty$ and equating the coefficients, one obtains the following superconvergence sum rules~\cite{Broniowski:2025ctl}:
\begin{align}\label{eq.sumrules_super}
    \frac{1}{\pi}\int_{t_\pi}^{\infty} \md t~t^{n-1} \operatorname{Im} \left(A,J,\Theta\right)(t)=0, \quad n=1,2,
\end{align}
that is, $\lim_{t\to\infty} t^n \times\left(A,J,\Theta\right)(t)=0$ due to the extra logarithmic suppression by the pQCD coupling~\cite{Broniowski:2025ctl}.
If we have a set of perfect representations of GFFs at any energy, these sum rules must be satisfied. However, without sufficient high-energy inputs, the dispersive results are only applicable at relatively low energies, and we cannot expect these sum rules to be fully saturated by the purely dispersive components, 
especially for the superconvergence sum rules, which certainly require the contribution of highly excited state poles to cancel each other out. 
Table~\ref{tab.sum rules} shows the degree of saturation of each sum rule from the $\pi\pi,K\bar{K}$ continua. Here, two integration limits are presented. 
The first is up to $\sqrt{t}=1.3~$GeV, which corresponds to the contribution from the $\pi \pi, K \bar{K}$ continua that are precisely known; above this energy, the $S$-wave two-channel approximation starts to break down, and a simple extrapolation without any obvious resonant structure is adopted for the $S$-wave phase shifts.
The second is at the two-nucleon threshold, $\sqrt{t}=2 m_N$. 
One indeed finds that the superconvergence sum rules require more high energy inputs to be saturated. 

\begin{table}[tb]
    \centering
    \caption{Degree to which the $\pi\pi,K\bar{K}$ continua saturate the nucleon GFF sum rules. Three GFFs, $(A,2J,\Theta/m_N)$, are normalized to unity at $t=0$ for easy illustration. Here, $n$ is the order of the weight in the superconvergence sum rules in Eq.~\eqref{eq.sumrules_super}, and the case of $n=0$ corresponds to Eq.~\eqref{eq.sumrules_norm}. The upper limit of the integration in the sum rules is specified by $\Lambda^2$ with $\Lambda=1.3$~GeV or $\Lambda=2m_N$.
     }
    \begin{tabular}{l | cc|cc|cc}
    \hline\hline
        & \multicolumn{2}{c}{$n=0$} & \multicolumn{2}{c}{$n=1$} & \multicolumn{2}{c}{$n=2$} \\ \hline
        $\Lambda$ & $1.3$~GeV & $2 m_N$ & $1.3$~GeV & $2 m_N$ & $1.3$~GeV & $2 m_N$ \\ \hline
        $A$ & $1.35$ & $1.77$ & $1.79$ & $2.58$ & $2.49$ & $4.04$ \\ \hline
        $2J$ & $1.19$ & $1.42$ & $1.43$ & $1.88$ & $1.91$ & $2.76$ \\ \hline
        $\Theta/m_N$ & $1.32$ & $2.60$ & $0.67$ & $3.71$ & $0.59$ & $8.09$ \\
    \hline\hline
    \end{tabular}
    \label{tab.sum rules}
\end{table}

Analogous to the superconvergence sum rule of nucleon GFFs~\eqref{eq.sumrules_super}, the meson GFFs $A^{\pi,K}$ also satisfy similar superconvergence sum rules~\cite{Broniowski:2024oyk}
\begin{align}\label{eq.super_SR_piK}
    \frac{1}{\pi} \int_{t_\pi}^{\infty} \mathrm{d} t~\operatorname{Im}A^{\pi,K}(t)=0,
\end{align}
that is, $\lim_{t\to\infty} t A^{\pi,K}(t)=0$.
The degrees of saturation of the above sum rules for the pion and kaon from the $\pi\pi,K\bar{K}$ continua are 1.86 and 0.77, respectively, when the integration upper limit is set to 2~GeV.

\section{$z$-expansion extrapolation of nucleon form factors}\label{app.z-exp}

It is known that the model-independent dispersive approach only predicts the low-energy behavior of the GFFs. 
However, to obtain the spatial distributions, we must extrapolate the GFFs up to $t\to -\infty$ while maintaining reasonable asymptotic behavior.
Unfortunately, the extrapolation cannot be determined in a model-independent way in practice. 
The widely used multipole model has been proven to be highly model-dependent in various analyses, see e.g.~\cite{Horbatsch:2015qda}.
Here, we perform a $z$-expansion extrapolation~\cite{Hill:2010yb} for the nucleon GFFs using the following expression,
\begin{align}\label{eq.z-exp}
    F^N\!\left(Q^2\right)=\mathcal{F}^N(z)=\sum_{k=0}^{k_{\max }} a_k z^k\ ,\quad z\left(t, t_{\mathrm{cut}}, t_0\right)=\frac{\sqrt{t_{\mathrm{cut}}-t}-\sqrt{t_{\mathrm{cut}}-t_0}}{\sqrt{t_{\mathrm{cut}}-t}+\sqrt{t_{\mathrm{cut}}-t_0}}\ ,
\end{align}
where $t=-Q^2$, $t_{\text{cut}}$ is set to be the two-pion threshold, i.e., $t_{\text{cut}}=t_\pi$, and $t_0$ is chosen to be its ``optimal'' value $t_0^{\text{opt}}\left(Q_{\max }\right)=t_{\text{cut}}\left(1-\sqrt{1+Q_{\max }^2 / t_{\text{cut}}}\right)$ to minimize the maximum value of $|z|$, with $Q_{\max}^2$ being the maximum $Q^2$ under consideration. 
The model-dependence of the extrapolation can be significantly suppressed by reaching a $k_{\max}$ region where the fitting results are insensitive to the choice of $k_{\max}$. 
To achieve such a $k_{\max}$ and ensure the QCD asymptotic behavior, we add constraints to the parameters based on the asymptotic behaviors of the GFFs in the limit $Q^2 \rightarrow \infty$. 
Here, we follow the method of Ref.~\cite{Wang:2024lrm}.

Using the definition in Eq.~\eqref{eq.z-exp}, in the limit $Q^2 \rightarrow \infty$, we have $z \rightarrow 1$ and $1-z\to 1/Q$, and the FF becomes
\begin{align}\label{eq.F-exp}
    \mathcal{F}^N(z)=\mathcal{F}^N(1)+\left.\sum_{n=1}^{k_{\max}} \frac{\md^n \mathcal{F}^N}{\md z^n}\right|_{z=1}(1-z)^n \sim \mathcal{F}^N(1)+\left.\sum_{n=1}^{k_{\max}} \frac{\md^n \mathcal{F}^N}{\md z^n}\right|_{z=1}\left(\frac{1}{Q^n}\right) .
\end{align}
If we assume that the FFs fall as positive powers of $1/Q$ at large $Q^2$, then we can write $\mathcal{F}^N(z \rightarrow 1) \propto 1 / Q^\ell$, where $\ell$ is an integer. 
Combining this with Eq.~\eqref{eq.F-exp}, we obtain the constraints for the coefficients $a_k$:
\begin{align}
    \mathcal{F}^N(1)= & \sum_{k=0}^{k_{\max}} a_k=0\ , \\
    \left.\frac{\md^n \mathcal{F}^N}{\md z^n}\right|_{z=1}= & \sum_{k=n}^{k_{\max}} \frac{k!}{(k-n)!} a_k=0, \quad n =1, \ldots, \ell-1\ .
\end{align}
In this work, we take $\ell=4$ for $A, J, \Theta$ and $\ell=6$ for $D$~\cite{Tong:2021ctu, Tong:2022zax}. 
We also set $k_{\max}=6$ and $k_{\max}=8$ for $A, J, \Theta$ and $D$, respectively, using $\left\{a_0, a_1, a_2\right\}$ as free parameters and leaving the rest of the coefficients $\left\{a_3, \ldots, a_6\right\}$ and $\left\{a_3, \ldots, a_8\right\}$ constrained. The corresponding fit parameters are listed in Table~\ref{tab.param}.
\begin{table}[tb]
    \caption{Parameters in the $z$-expansion extrapolation, cf. Eq.~\eqref{eq.z-exp}, obtained by fitting to the dispersive results of the four GFFs, $A,J,\Theta$ and $D$. The errors listed here are statistical only.}
    \centering
    \begin{tabular}{l | cccc}
    \hline\hline
        & $A$ & $J$ & $\Theta/m_N$ & $D$ \\ \hline
        $a_0$ & $0.568(1)$ & $0.264(0)$ & $0.079(1)$ & $-1.080(1)$ \\ \hline
        $a_1$ & $-1.208(4)$ & $-0.572(2)$ & $-0.638(3)$ & $4.202(7)$ \\ \hline
        $a_2$ & $-0.120(8)$ & $-0.054(4)$ & $1.426(9)$ & $-4.276(25)$ \\
    \hline\hline
    \end{tabular}
    \label{tab.param}
\end{table}





\end{appendices}

\bibliography{refs}

\end{document}